\documentclass[structabstract]{aa}
\usepackage{natbib}
\bibpunct{(}{)}{;}{a}{}{,}
\usepackage{graphicx}

\newcommand{\kms}{\ensuremath{{\rm km\,s^{-1}}}}                     
\newcommand{\micron}{\ensuremath{\mu\mathrm{m}}}                     

\begin{document}

\title{VLTI/AMBER spectro-interferometry of the Herbig Be star MWC~297 with spectral resolution 12\,000
\thanks{
Based on observations made with ESO telescopes at Paranal Observatory under programme ID 081.D-0230(A).} }

\author{G.~Weigelt         \inst{1}
  \and V.~P.~Grinin        \inst{2,3}
  \and J.~H.~Groh          \inst{1}
  \and K.-H.~Hofmann       \inst{1}
  \and S.~Kraus            \inst{4}
  \and A.~S.~Miroshnichenko\inst{5}
  \and D.~Schertl          \inst{1}
  \and L.~V.~Tambovtseva   \inst{2}
  \and M.~Benisty          \inst{6}
  \and T.~Driebe           \inst{7}
  \and S.~Lagarde          \inst{8}
  \and F.~Malbet           \inst{9}
  \and A.~Meilland         \inst{1,8}
  \and R.~Petrov           \inst{8}
  \and E.~Tatulli          \inst{9}
  }

\offprints{G.~Weigelt\\ email: \texttt{weigelt@mpifr.de}}

\institute{Max-Planck-Institut f\"ur Radioastronomie, Auf dem H\"ugel 69, D-53121 Bonn, Germany
\and Pulkovo Observatory of RAS, Pulkovskoe shosse 65, St. Petersburg 196140, Russia
\and V.V. Sobolev Astronomical Institute, St. Petersburg University, St. Petersburg, Russia
\and Department of Astronomy, University of Michigan, 500 Church Street, Ann Arbor, MI 48109, USA
\and University of North Carolina at Greensboro, P.O. Box 26170, Greensboro, NC 27402, USA
\and INAF-Osservatorio Astrofisico di Arcetri, Istituto Nazionale di  Astrofisica, Largo E.~Fermi 5, I-50125 Firenze, Italy
\and  Deutsches Zentrum f\"ur Luft- und Raumfahrt e.V., K\"onigswinterer Str. 522-524, D-53227 Bonn, Germany
\and Laboratoire H. Fizeau, University of Nice Sophia Antipolis, CNRS, OCA, Parc Valrose, 06108 Nice, France
\and Laboratoire d'Astrophysique de Grenoble, UMR 5571 Universit\'e Joseph  Fourier/CNRS, BP 53, F-38041 Grenoble Cedex 9, France
 }

\authorrunning{G.~Weigelt et al.\ }
\titlerunning{Spectro-interferometry of MWC~297 with  VLTI/AMBER}

\date{Received  / Accepted }

\abstract
{Circumstellar disks and outflows play a fundamental role in star formation. Infrared spectro-interferometry allows the inner accretion-ejection
region to be resolved. }
{We  study  the disk and  Br$\gamma$-emitting region of \object{MWC 297} with high spatial and spectral resolution and compare our observations with
disk-wind models.
    }
{We measured interferometric visibilities, wavelength-differential phases, and closure phases of MWC~297 with a spectral resolution of 12\,000. To
interpret our MWC~297 observations, we employed disk-wind models.
    }
{The measured continuum visibilities confirm previous results that the continuum-emitting region of MWC~297 is remarkably compact. We derive a
continuum ring-fit radius of $\sim$2.2~mas ($\sim$0.56~AU at a distance of 250~pc), which is $\sim$5.4 times smaller than the 3~AU dust sublimation
radius expected for silicate grains (in the absence of radiation-shielding material). The strongly wavelength-dependent and asymmetric
Br$\gamma$-emitting region is more extended ($\sim$2.7\,times)  than the continuum-emitting region. At the center of the Br$\gamma$ line, we derive a
Gaussian fit radius of $\sim$6.3~mas HWHM ($\sim$1.6~AU). To interpret the observations, we employ a magneto-centrifugally driven disk-wind model
consisting of an {\it accretion disk}, which emits the observed continuum radiation, and a {\it disk wind}, which emits the Br$\gamma$ line. The
calculated wavelength-dependent model intensity distributions and Br$\gamma$ line profiles are compared with the observations (i.e., $K$-band
spectrum, visibilities, differential phases, and closure phases). The closest fitting model predicts a continuum-emitting disk with an inner radius
of $\sim$0.3~AU and a disk wind ejection region with an inner radius of $\sim$0.5~AU ($\sim$17.5 stellar radii). We obtain a disk-wind half-opening
angle (the angle between the rotation axis and the innermost streamline of the disk wind) of $\sim$80$\degr$, which is larger than in T Tau models,
and a disk inclination angle of $\sim$20$\degr$ (i.e., almost pole-on).
      }
{Our observations with a spectral resolution of 12\,000 allow us to  study the AU-scale environment of MWC~297 in $\sim$10 different spectral
channels across the Br$\gamma$ emission line. We show that the $K$-band flux, visibilities, and remarkably strong phases can be explained by the
employed magneto-centrifugally driven disk wind model. }


\maketitle

\section{Introduction \label{intro}}

Circumstellar disks and outflows play  a fundamental role in star formation. Testing theoretical accretion-ejection models is very challenging since
important processes occur on AU and sub-AU scales, which cannot be resolved by conventional imaging. However, infrared spectro-interferometric
observations with milli-arcsecond spatial resolution and high spectral resolution allow unprecedented studies of the inner accretion region, which
hosts fascinating astrophysical objects such as inner gaseous accretion disks, dust disks, disk winds, jets, etc. Interferometry in the near- and
mid-infrared is able to probe the gas and dust distribution on these scales and constrain their physical properties. These observations allow us to
study the nature of the accretion process and the launching of jets and winds.

With a spectral type of B1.5V, a mass of $\sim$10~$M_{\sun}$, and a distance of $250\pm 50$~pc \citep{dre97}, MWC~297 is one of the nearest massive
young stars.  The geometry of the disk of MWC~297 has been investigated by various techniques at near-infrared
(\citealt{mil01,eis04,mon06,mal07,ack08,kra08a}), mid-infrared (\citealt{ack08}), and radio wavelengths \citep{man97,dre97}. Interestingly, several
interferometric observations of MWC~297 (\citealt{EISNER04,MONNIER05,mal07,ack08,kra08a}) have shown that the compact characteristic size of the
near-infrared continuum-emitting  region is several times smaller than the theoretically expected $\sim$3~AU radius of the dust sublimation rim (the
expected radius if there is no shielding of the stellar radiation; \citealt{MONNIER05}).

The estimates of the inclination angle  of MWC~297's circumstellar  disk have been highly controversial. \citet{dre97} argued for an almost edge-on
disk orientation. Moderate inclination angles of the circumstellar disk of MWC~297 were found with infrared interferometric observations. For
instance, \citet{mal07} obtained VLTI/AMBER observations with a medium spectral resolution of 1500 and measured the wavelength dependence of the
visibility across the Br$\gamma$ line. \citet{mal07} interpreted their spectro-interferometric measurements of MWC~297 with an anisotropic
stellar-wind model \citep{Stee94,Stee95} and found an inclination of $\sim$20$\degr$. \citet{ack08} obtained both near-infrared AMBER $H$- and
$K$-band observations and mid-infrared MIDI observations and concluded that the inclination is smaller than $40\degr$.

In this paper, we present the first spectro-interferometric measurements of the inner circumstellar environment of MWC~297 with AMBER's  high
spectral resolution mode ($R$\,=\,12\,000). This high spectral resolution allows us to spectrally resolve the Br$\gamma$ emission line and measure
the interferometric observables (spectrum, visibilities, wavelength-differential phases, and closure phases) in $\sim$10 different spectral channels
across the line. In Sect. 2, we present the AMBER observations. In Sect. 3, we derive characteristic continuum and Br$\gamma$ line sizes of the
circumstellar environment using simple geometric models. In Sect. 4, we present radiative transfer modeling of the disk and disk wind and compare the
$K$-band flux, visibilities, and phases with the model predictions in the continuum and several spectral channels across the Br$\gamma$ emission
line. In the Appendix, we discuss the disk-wind modeling, spectroscopy of MWC~297, wavelength calibration, and the use of AMBER's Beam Commutation
Device.


\section{Interferometric VLTI/AMBER/FINITO observations with spectral resolution of 12\,000}\label{sect_obs}

\begin{figure}
\hspace*{3mm}
\vspace*{-20mm}
\includegraphics[width= 8.2cm, angle =0]{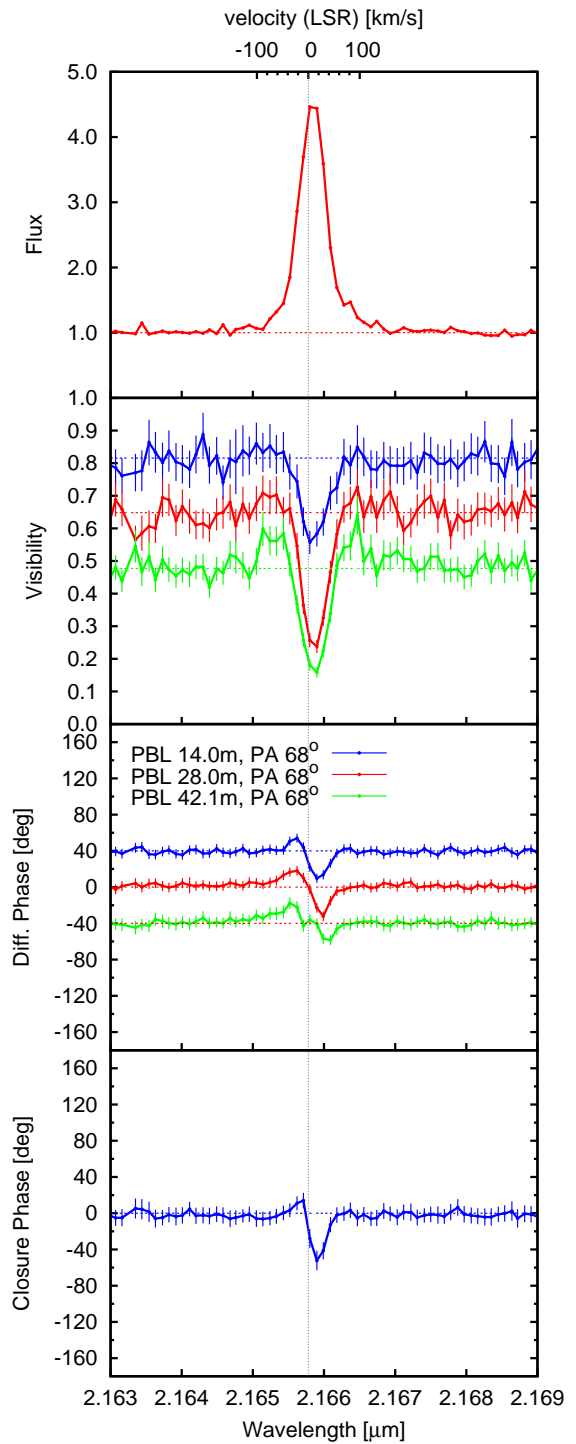}

\vspace*{10mm}
\caption{\label{obs_fig1} AMBER observations of MWC~297 with spectral resolution of 12\,000: (from  top to bottom) wavelength
dependence of flux, visibilities, wavelength-differential phases (for better visibility, the differential phases of the shortest and longest
baselines are shifted by +40 and $-40\degr$, respectively), and closure phase observed at projected baselines of 14.0, 28.0, and 42.1~m along the
position angle (PA) of $68.0\degr$ on the sky. The wavelength scale at the bottom is the directly observed one, i.e. without heliocentric or LSR
correction (see text). However, the radial velocity scale at the top, gives spectrum, visibilities, and phases as a function of the Br$\gamma$
Doppler shift in the Local Standard of Rest (LSR) frame. The dashed vertical line indicates the centroid Br$\gamma$ vacuum wavelength (2.1661 $\mu$m)
in the LSR frame.}

\end{figure}

We observed MWC\,297 on 6 April 2008 with  ESO's Very Large Telescope Interferometer (VLTI) and its AMBER beam combiner instrument  in the course of
the AMBER-GTO program 081.D-0230(A). AMBER \citep{pet07} allows interferometric three-telescope observations with low, medium, or high spectral
resolution. The observations are described in Table~\ref{obs}. For these observations, the E0-G0-H0 array of the 1.8~m auxiliary telescopes (ATs) and
AMBER's high resolution mode (HR mode; $R$~=~12\,000) were employed. To obtain interferograms with a high signal-to-noise ratio (SNR), we used the
fringe tracker FINITO and a detector integration time of 8.0~seconds per interferogram. With excellent atmospheric conditions (seeing
0.4--0.6\arcsec), the FINITO instrument in the co-phasing mode allowed us to stabilize the atmospheric and mechanically induced fringe motion. FINITO
records temporally modulated $H$-band interferograms with a high frequency, from which the atmospheric phase shift is determined and then compensated
for using the VLTI delay lines.

\begin{table*}[t]
\caption{\label{obs} Log of the VLTI/AMBER/FINITO observations  of MWC~297.} \vspace*{3mm} \label{tab:observations} \centering
\begin{tabular}{lll ccccccccc}
  \hline\hline
Date&\multicolumn{2}{c}{Time [UT]} &AT array &Spectral &Wavelength &DIT$^a$ &$N_{\rm t}^b$ &Calibrator &$N_{\rm c}^c$ &Uniform-disk\\
    &Start      &End               &         &mode     &Rrange     &        &              &           &              &diameter of the \\
    &           &                  &         &         &[$\mu$m]   &[s]     &              &           &              &calibrator [mas]\\
  \noalign{\smallskip}
  \hline
  \noalign{\smallskip}
  2008 Apr. 06  &08:14  & 08:57            &E0-G0-H0      &HR-K-F$^d$ &2.147--2.194  &8        &200          &HD\,175\,583  &80 & 1.02$^e$\\
  \noalign{\smallskip}
  \hline
\end{tabular}

\begin{flushleft}
 \hspace{0mm}Notes~--~$^{a}$ Detector integration time per interferogram;
 $^{b}$ Number of MWC~297 interferograms;
 $^{c}$ Number of calibrator interferograms;
 $^{d}$ HR mode in the $K$ band using the fringe tracker FINITO;
 $^{e}$ Uniform-disk diameter taken from \citet{mer2005}.\\
\end{flushleft}
\vspace*{-3mm}

\end{table*}

For data reduction, we used AMBER's {\it amdlib} data reduction software (release 2.2), which employs the P2VM algorithm \citep{tat07} to derive
wavelength-dependent visibilities, wavelength-differential phases, and closure phases. In addition, the spectrum can be extracted as an additional
observable.

Along with the science observation of MWC~297, we observed the interferometric calibrator star HD\,175\,583, which allowed us to calibrate the
atmospheric transfer function. We recorded 100 target interferograms (plus 40 calibrator interferograms) without AMBER's beam commutation device
(BCD~OUT data) and 100 target interferograms plus 40 calibrator interferograms with the BCD (BCD~IN data). The BCD allows the study of systematic
instrumental effects (see Appendix~\ref{bcd}). A fraction of the interferograms were of lower quality because of the  lower  FINITO performance.
Therefore, we selected  the 50\% of the BCD~OUT data and 50\% of the BCD~IN data with the highest fringe SNR among both the target and the calibrator
interferograms to improve the visibility calibration \citep{tat07}. From this data, we derived the spectrum, visibilities, wavelength-differential
phases, and closure phases, observed along the position angle (PA) of 68.0$\degr$ on the sky (see Fig.~\ref{obs_fig1}). The differential phases and
closure phases derived from both independent data sets (BCD~OUT and BCD~IN) are shown in Fig.~\ref{bcdfig} of Appendix \ref{bcd}. Subtraction of the
closure phases obtained with and without BCD cancels potential instrumental phase drifts. Figure~\ref{bcdfig2} presents the derived closure phase and
a fit of a constant to the continuum closure phases. The obtained averaged continuum (2.147--2.194~$\mu$m) closure phase is $-1.28 \pm 0.42\degr$.

With the high spectral resolution of $R=12\,000$, it is possible to spectrally resolve the Br$\gamma$ line and measure the visibilities and phases in
$\sim$10 different spectral channels across the Doppler-broadened Br$\gamma$ line. The visibilities show a strong wavelength dependence of the size
of the line-emitting region. The wavelength-differential phases and the closure phase indicate  remarkably strong photocenter shifts and asymmetries,
respectively,  of the circumstellar environment in many spectral channels across the  Br$\gamma$ line. We  show in Sect. \ref{diskwind} that this
strong wavelength dependence and the large values of the phases can be explained by disk-wind models.

The wavelength calibration of the AMBER data in Fig. \ref{obs_fig1} was accomplished using the many telluric lines present in the region
2.15--2.19~$\micron$, as described in the Appendix \ref{wavecal}. We estimate an uncertainty in the wavelength calibration of $\sim$3~$\kms$. The
spectrum in Fig.~\ref{obs_fig1} (top) was normalized to the nearby continuum. The wavelength scale at the bottom is shown without heliocentric or
Local Standard of Rest (LSR) correction.  However, the radial velocity scale at the top of Fig.~\ref{obs_fig1} is calibrated to give spectrum,
visibilities, and phases as a function of the Br$\gamma$ Doppler shift in the Local Standard of Rest (LSR) frame ($16.5~\kms$ has to be subtracted to
convert the LSR values to the heliocentric frame).

The obtained  Br$\gamma$ line profile has an equivalent width  of $-17.2$~{\AA}. Using AMBER's HR mode, we were able to resolve the  Br$\gamma$ line
and determined a FWHM of $\sim$60~\kms. By performing a Gaussian fitting of the line profile, a centroid LSR velocity of $13.8 \pm 3~\kms$ was
obtained. No P-Cygni absorption or double-peaked emission is seen. The FWHM of the Br$\gamma$ line is compatible with the value of $53~\kms$ found by
\citet{dre97} for Br$\alpha$. On the other hand, the centroid velocity of Br$\gamma$ is more similar to the one measured for H$\alpha$ and H$\beta$
(16 and $19~\kms$, respectively; \citealt{dre97}) than to that of Br$\alpha$ ($3~\kms$, \citealt{dre97}).


%
\begin{figure*}[]
\begin{centering}
\vspace{-4mm}
\includegraphics[width=154mm, angle =-0]{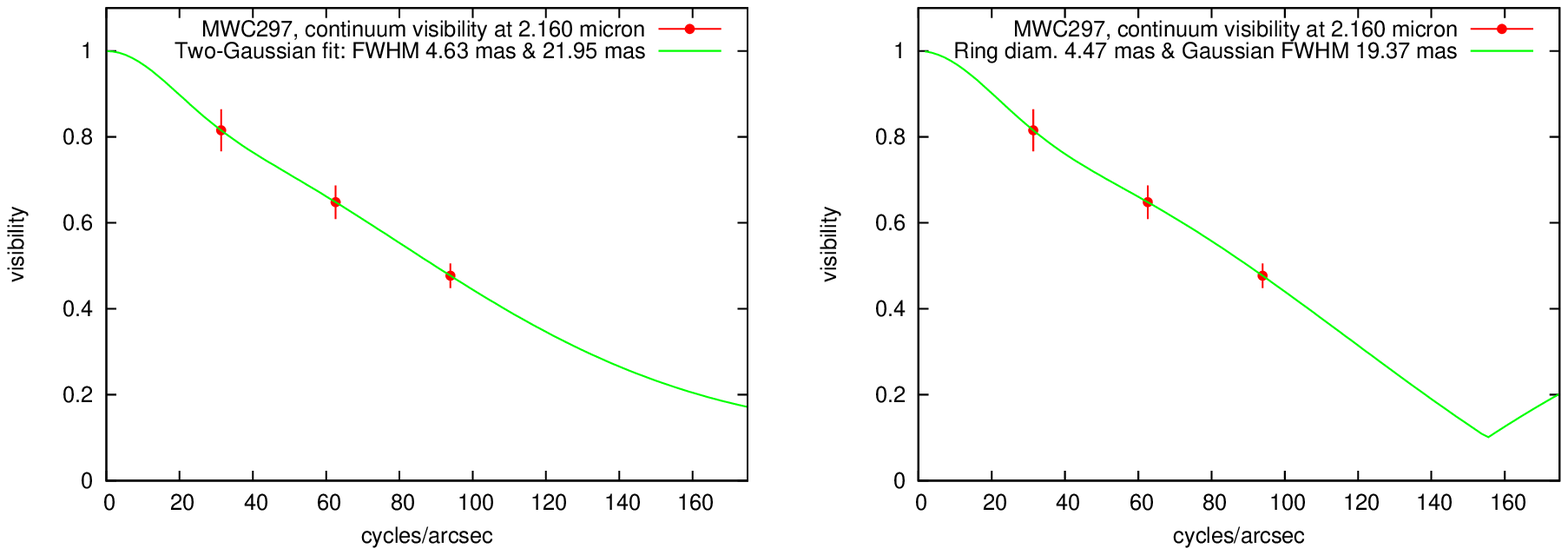}
\vspace{-160mm}
\caption{\label{continuum fit}{\it Left:} Fit of a two-component Gaussian plus a point source with a 10\% flux contribution to the
wavelength-averaged (2.147--2.194\,$\mu$m)  $K$-band continuum  visibilities near Br$\gamma$. {\it Right:} Fit of a compact ring plus Gaussian halo
plus a point source with a 10\% flux contribution to the $K$-band continuum visibilities. }
\end{centering}
\end{figure*}

%
\begin{figure*}[]
\begin{centering}
\vspace{-4mm}
\hspace{-5mm}
\includegraphics[width=112mm, angle =-90]{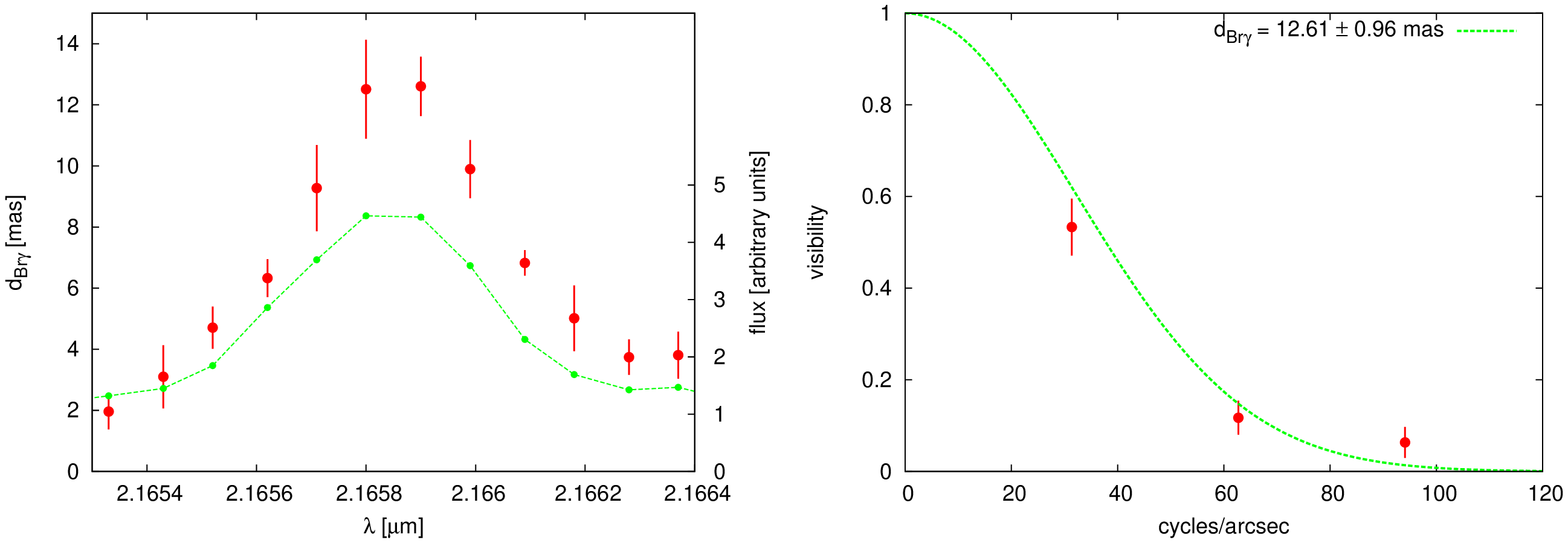}
\vspace{-56mm}
\caption{\label{brg} {\it Left:} Gaussian FWHM fit diameters of the Br$\gamma$ line-emitting region for several individual spectral channels  (in the
center and the line wings) across the Br$\gamma$ emission line (red bullets). The diameters are derived from continuum-compensated visibilities.
Therefore, they are the diameters of the Br$\gamma$ line-emitting region and not of the emission region of both line plus continuum. The AMBER
spectrum of MWC~297 is shown for comparison (green curve). {\it Right:} Gaussian fit to the continuum-compensated AMBER  visibilities  at the peak
intensity of the Br$\gamma$ emission line. }
\end{centering}
\end{figure*}

\begin{table*}[t]
\caption{\label{fit} Geometric model fit radii of MWC~297 in the continuum and the center of the Br$\gamma$ emission line.} \vspace*{3mm}
\label{tab:fits} \centering
\begin{tabular}{llllll}
  \hline\hline
Wavelength             &component C1            &diameter of C1      &component C2  &diameter of C2  &flux ratio\\
range                  &(dominant compact core) &                    &(halo)        &                &(C1/C2)\\
  \noalign{\smallskip}
  \hline
  \noalign{\smallskip}
Continuum$^{a}$        &Gaussian                &$4.63 \pm 0.21$~mas  &Gaussian     &$\ga 22$~mas $^{c}$ &4.6  \\
Continuum$^{a}$        &Ring (20\%)$^{b}$       &$4.47 \pm 0.20$~mas  &Gaussian     &$\ga 19$~mas $^{c}$ &3.6  \\
Br$\gamma$ center        &Gaussian                &$12.6 \pm 0.75$~mas $^{d}$  &-            &-                &-    \\

  \noalign{\smallskip}
  \hline
\end{tabular}

\begin{flushleft}
 \hspace{0mm}Notes~--~
$^{a}$ average over continuum wavelengths between 2.147 and 2.194\,$\mu$m (except the line region). $^{b}$ Inner radius of a ring model with a ring
width of 20\% of the inner ring radius. $^{c}$ Due to the lack of data at small spatial frequencies, the size of the extended halo component is not
well constrained by our measurements. Therefore, we can only give a rough estimate: $22(19) \pm 5$~mas or larger, but smaller than the AMBER AT FOV
of 250~mas. $^{d}$ Continuum-compensated diameter (see text).
 \\
\end{flushleft}
\vspace*{-3mm}

\end{table*}


\section{Comparison of the observations with geometric models}\label{geometric}


\subsection{Characteristic size  of the continuum-emitting region}

To measure the characteristic size  of the circumstellar environment in the $K$-band continuum,  we fitted geometric models to the visibilities
obtained along the PA of $68.0\degr$ on the sky (see Table~\ref{fit}). As pointed out by \citet{ack08}, who observed MWC~297 with the same VLTI
E0-G0-H0 array, one-component uniform-disk, ring, and Gaussian disk models are unable to fit the shape of the continuum visibility. Therefore, we
fitted a two-component Gaussian plus a point source with a 10\% flux contribution (derived from the SED; as in \citealt{ack08}) to the
wavelength-averaged continuum visibilities. Our observations are averages over the continuum wavelengths between 2.147 and 2.194\,$\mu$m (except the
line region) because of the narrower wavelength range of AMBER's HR mode.

Figure~\ref{continuum fit} (left) shows that the fitted two-component Gaussian consists of a dominant compact Gaussian with a FWHM diameter of $4.63
\pm 0.21$~mas (corresponding to $1.16 \pm 0.05$~AU at a distance of 250 pc) in addition to an extended Gaussian halo with a FWHM diameter of
$\sim$22~mas (see Table~\ref{fit}). Owing to the lack of data at low spatial frequencies, the size and shape of the extended halo component is not
well constrained. The flux ratio of the dominant compact component to the extended halo component is 4.6. Our FWHM diameter of $\sim$4.6~mas obtained
for the dominant compact Gaussian component is in good agreement with the diameter of $\sim$4.3~mas reported by \citet{ack08}.

To verify MWC~297's location in the size-luminosity relation (where ring fits are used; \citealt{MONNIER05,kra08a}), we  determined its ring-fit
radius for a ring thickness of 20\% of the radius. Figure~\ref{continuum fit} (right) shows the fit obtained of a compact ring, an extended Gaussian
halo, and a 10\% point source to the wavelength-averaged (2.147--2.194\,$\mu$m) continuum visibilities. The fitted  FWHM diameter (inner ring
diameter) of the dominant compact model ring  is  $4.47 \pm 0.20$~mas (corresponding to $1.12 \pm 0.05$~AU for a distance of 250 pc; see
Table~\ref{fit}).

The measured continuum ring radius of 0.56~AU is $\sim$5.4 times smaller than the $\sim$3~AU dust sublimation radius of MWC~297 predicted by the
size-luminosity relation, if there is no shielding of the stellar radiation (\citealt{MONNIER05,kra08a}). This surprisingly compact size was already
discussed by several other authors (\citealt{eis04,MONNIER05,mon06,mal07,kra08a,ack08}). The correct interpretation of the compact continuum-emitted
region of MWC~297 and several other Herbig Be stars remains disputed. Possible explanations include the absorption of the stellar light by an inner
gas disk allowing dust to survive closer to the star (e.g., \citealt{mil01,MONNIER05,mal07}), emission from an optically thick inner gas disk (e.g.,
\citealt{MONNIER05,kra08b}), and emission from special types of dust grains that can survive at higher temperatures (e.g., \citealt{ben10}).


\subsection{The characteristic size  of the Br$\gamma$-emitting region}

The high spectral resolution of the AMBER data allows us to measure the characteristic size of the Br$\gamma$ line-emitting region for several
individual spectral channels across the Br$\gamma$ emission line (see Fig.~\ref{brg} left).  These diameters were determined by fitting Gaussians to
the continuum-compensated  Br$\gamma$ visibilities (which are the visibilities free from contributions from the underlying continuum within the line
region). The right panel of Fig.~\ref{brg} shows such a Gaussian fit for the central wavelength of the Br$\gamma$ emission line.

The continuum-compensated visibilities required for the size determination of the line-emitting region discussed above were calculated in the
following way. Within the  wavelength region of Br$\gamma$, the measured visibility has two constituents, a pure line-emitting component and a
continuum-emitting one, the second of which includes emission of both the circumstellar environment and the unresolved star. The emission line
visibility $V_{{\rm Br}\gamma}$ can be written as \citep{wei07}


\begin{eqnarray}\label{eq_brgvis1}
& & F_{{\rm Br}\gamma}V_{{\rm Br}\gamma} = \nonumber \\
& & = \sqrt{ |F_{\rm tot}V_{\rm tot}|^2 + |F_{\rm c}V_{\rm c}|^2 - 2\,F_{\rm tot}V_{\rm tot}\,F_{\rm c}V_{\rm c}\cdot cos{\Phi} },
\end{eqnarray}

\noindent where $V_{\rm tot}$ ($F_{\rm tot}$) denotes the measured total visibility (flux) in the  ${\rm Br}\gamma$ line, $V_{\rm c}$ ($F_{\rm c}$)
is the measured visibility (flux) in the continuum, and $\Phi$ describes the measured wavelength-differential phase within the ${\rm Br}\gamma$ line.
>From the AMBER spectrum  (see Fig.~\ref{obs_fig1}), we find $F_{\rm tot}=4.5$ and $F_{{\rm Br}\gamma}=3.5\times F_{\rm c}$ at the emission line
center.

We derived a large continuum-compensated radius (HWHM) of $6.3 \pm 0.4$~mas ($1.6 \pm 0.1$~AU) for the line-emitting region at the peak wavelength of
the emission line, which is $\sim$3 times larger than the 0.56~AU HWHM radius of the compact continuum-emitting Gaussian component. The smaller
line-continuum size ratio obtained by \citet{kra08a} can be explained by the lower spectral resolution of $R = 1500$, which leads to an averaging
over different diameters corresponding to different wavelengths. Figure~\ref{brg} also shows that the size of the line-emitting region is smaller in
the wings of the Br$\gamma$ emission line than at the line center. This wavelength dependence is one of several observational results that have to be
explained by models, such as the  disk-wind model  presented in the next section.


\section{Comparison of the observations with a model of a magneto-centrifugally driven disk wind}\label{diskwind}


\begin{figure}[]\label{sk}
\begin{centering}
\vspace{5mm}
\includegraphics[width=80mm, angle =0]{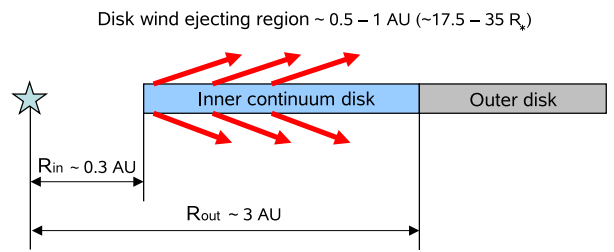}
\vspace{-0mm}
\caption{Sketch of the MWC~297 disk-wind model adopted in this paper.}
\end{centering}
\end{figure}


We now present a disk-wind model and compare the observed spectrum, visibilities, and phases of MWC~297 with the model predictions. In this disk-wind
model, mass ejection is driven by magneto-centrifugal forces. The theory of these winds was originally proposed by \citet{BP82} and has been
elaborated by many authors, mainly for T~Tauri stars (see \citealt{kon00}, \citealt{pud05}, \citealt{fer07}, and references therein). An important
property of the launching mechanism in magneto-centrifugally driven disk winds is the topology of the magnetic field in the disk: magnetic field
lines in the wind-launching region are inclined with respect to the disk plane by less than $60\degr$ (Blandford \& Payne 1982). The radiation
pressure of hot stars probably changes the trajectories of the gas streams, leading to a flatter disk wind \citep{dek95,DREW98,eve01}.
Figure~\ref{sk} shows a sketch of the disk-wind model that we consider. It assumes that both an accretion disk and a disk wind contribute to the
observed emission.

\subsection{Geometry and kinematics of the disk  wind}
To model the extended disk wind, we follow an approach used by \citet{SV93} in their studies of  disk winds in cataclysmic variables. This approach
provides a simple parametrization of a disk wind, which has geometrical and kinematical properties similar to those used by \citet{KHS06} for T Tauri
stars (see  Fig.~\ref{sketch} in the Appendix). Here, we use a similar description of the kinematical parameters of the wind where according to the
conservation of angular momentum, the tangential velocity $u$ decreases along a streamline from the Keplerian velocity at the base of streamline. The
radial velocity $v$ increases along the streamline from the initial value $v_0$ to $v_{\infty}$ (see Eqs.~\ref{velu} and \ref{velv} and
Fig.~\ref{vel} in  Appendix A), where $v_0$ and $v_{\infty}$ are the model parameters.

The local mass-loss rate per unit area of the disk, $\dot m_w$, is a function of the cylindric radius $\omega$. To describe $\dot m_{w}$, we use a
simple power-law
\begin{equation}\dot m_{w}(\omega) \sim \omega^{-\gamma}\,,
\end{equation}
where $\gamma$ is a free parameter. The other model parameter is the total mass-loss rate
\begin{equation}
\dot M_{w} = 2\int_{\omega_{1}}^{\omega_N}\dot m_{w}(\omega)\,2\,\pi\,\omega\,d\omega ,
\end{equation}
where $\omega_{1}$ and $\omega_N$ are the inner and outer boundaries of the disk-wind ejection region, and the factor two accounts for the mass loss
from both sides of the disk surface.

\subsection{The parameters of the accretion disk}
In classical accretion-disk models, the surface density increases toward the star as
\begin{equation}
\Sigma(\omega) \sim \omega^{-p}\,,
\end{equation}
where $p$ = 1--1.5 \citep[see, e.g.,][]{CHS00}. At a mass accretion rate of $\dot M_{acc} \geq 10^{-7} M_\odot{\rm yr}^{-1}$, this disk is optically
thick (due to the gas opacity) between the dust evaporation radius and either the stellar magnetosphere \citep{MUZEROLLE04} or the stellar surface.
For MWC~297, the value of $\dot M_{acc}$ is fairly uncertain \citep[see ][] {mal07}. If the surface density in the accretion disk of MWC~297 changes
as $\omega^{-1}$, then this disk will be optically thick up to the stellar surface. However, if the accretion disk extended to the innermost vicinity
of the star, the disk luminosity at near-infrared wavelengths would exceed the observed one. This is the main reason why an inner disk gap is
required.

As in  \citet{ack08}, we used a blackbody approximation with a power-law distribution of the disk temperature $T_d$
\begin{equation}
T_d(R) = T_d(R_{in})(R/R_{in})^\alpha\,,
\end{equation}
where $R_{in}$ is the inner radius of the disk, and $\alpha$ is a parameter. The parameter $\alpha$ is equal to $-3/4$ in the two important special
cases when a flat disk is just heated by the stellar radiation (passive disk; \citealt{AS86}) or radiates due to the viscous dissipation
\citep{Lyn74}. However, this disk (as well as other simple power-law  disk models) fails to reproduce the shape of the observed continuum
visibilities \citep{ack08}. More sophisticated models can be obtained with two-component temperature-gradient models with slightly different
temperature gradients (see below).

%
\subsection{Model calculations}\label{modelcalculations}

For simplicity, it is assumed that the disk wind contains only hydrogen atoms. For the calculations of the ionization state and the number densities
of the atomic levels, we adopted the numerical codes developed by \citet{GM90} and \citet{TAMBOVTSEVA01} for moving media. These codes are based on
the \citet{VVS} approximation in combination with the exact integration of the line intensities (see Appendix A). This method takes into account the
radiative coupling in the local environment of each point caused by multiple scattering.

The stellar radiation of MWC~297 is described by a model atmosphere \citep{KURUCZ79} with an effective temperature $T_{\rm eff} = 25\,000$~K, log g =
4, and the stellar radius $R_*$ = 6.1\,$R_\odot$ as in \citet{dre97}. The calculations were performed for an isothermal wind. We adopted electron
temperature $T_e$ values in the range  8000--10\,000~K, which are typical of the disk winds of young stars (see, e.g., Kurosawa et al. 2006). In our
models, we adopted the  fixed parameters  stellar mass $M_*$ = 10~$M_\odot$ and a distance of 250\,pc \citep{dre97}.


\begin{figure}[]
\begin{centering}
\includegraphics[width= 77mm, angle =-0]{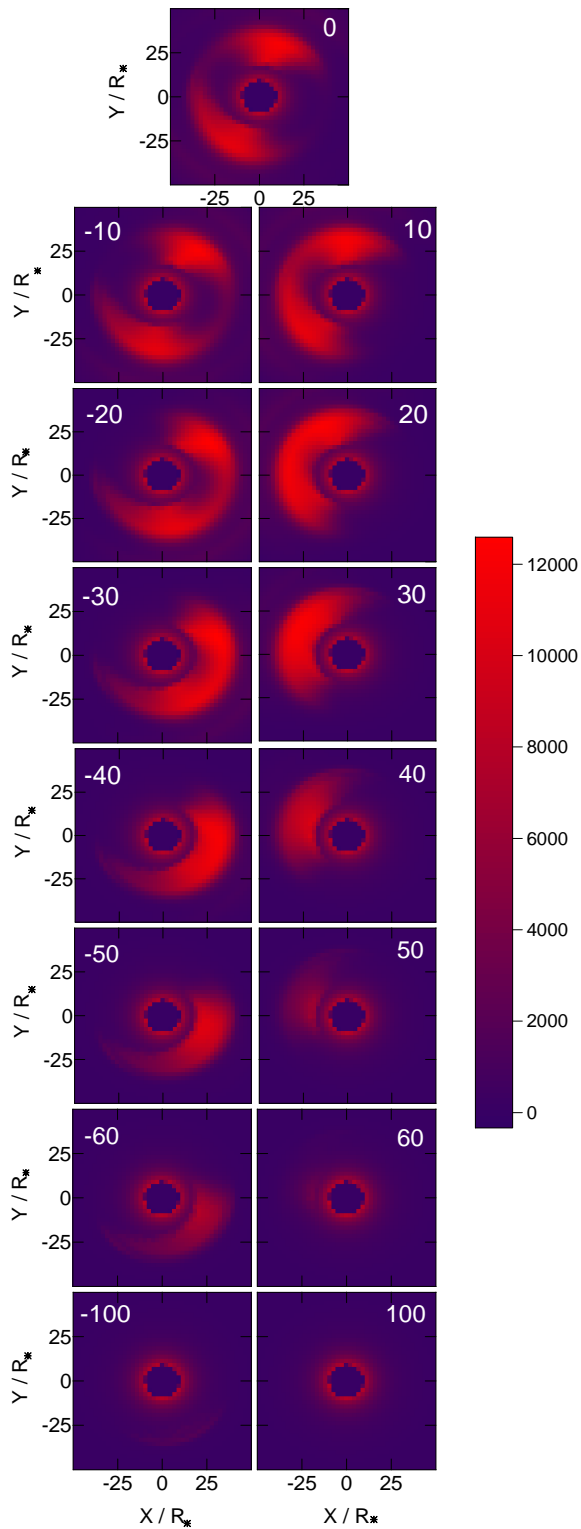}
\caption{\label{lmap} Intensity distributions of our best-fit disk-wind model 5 (i.e., intensity distribution of the continuum disk plus the disk
wind; the central star is not shown; see Tables~\ref{tab3} and \ref{modeltab}) at the center of the Br$\gamma$ line ($v=0~\kms$) and at 14 other
velocities (the  labels give the velocity in $\kms$). For the calculation of the model images in this figure, a clockwise motion of the disk wind was
assumed. Therefore, in the blue-shifted images (left panels), mainly the disk regions on the right hand side of the star are bright. The radius of
the inner edge of the disk wind ejection region (i.e., radius of the inner hole) is $\omega_1$ = 17.5~$R_*$ ($\sim$0.3~AU). The inclination angle
(angle between the polar axis and the viewing direction) of the model is $i$ = 20$\degr$ (i.e., almost pole-on). The colors represent the intensity
in erg ster$^{-1}$\,s$^{-1}$\,\AA$^{-1}$\,cm$^{-2}$.  In these images, AMBER's spectral resolution of 12\,000 was modeled, as described in Sect.
\ref{ip}.}
\end{centering}
\end{figure}


Figure~\ref{lmap} presents examples of the intensity distributions (projected on the plane of the sky) of our best-fit continuum-disk and disk-wind
model number 5 (see Tables \ref{modeltab} and \ref{tab3}). The inclination angle of the model (i.e., angle between the polar axis and the viewing
direction) is $i$ = 20$\degr$, i.e., almost pole-on. Model images  at 15 different wavelengths (Doppler shifts) across the Br$\gamma$ line profile
are shown. The images are highly asymmetric and wavelength-dependent. Their wavelength-dependent photocenter shift (with respect to the continuum
image) and asymmetric shape change across the line, resulting in wavelength-dependent visibilities, differential phases, and closure phases. This
asymmetry of the model images is the result of the Doppler shift of the radiation due to the gas motion in the emitting region. The main part of the
Br$\gamma$ line emission is formed near the base of the disk wind where the dominant motion is  Keplerian rotation. For the calculation of the model
images shown in Fig.~\ref{lmap}, a clockwise motion of the disk wind was assumed. When comparing with the observations, we calculated the model
images for both clockwise and anti-clockwise motion.


\begin{figure}[h]
\centering
\includegraphics[width= 4.9cm, angle =-90]{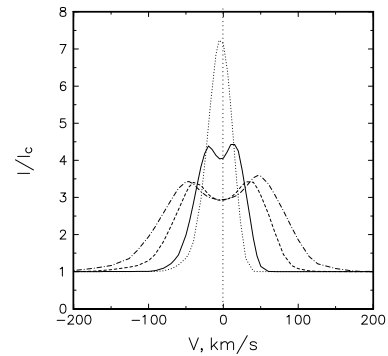}
\caption{\label{BRG} Normalized Br$\gamma$ line profiles of the disk-wind model 5 from Table A.1 at the inclination angles $i$ = 10$\degr$ (dotted
line), 20$\degr$ (solid), 40$\degr$ (dashed), and 60$\degr$ (dashed-dotted; high spectral resolution, not degraded to the spectral resolution of
AMBER).}
\end{figure}


Examples of  theoretical Br$\gamma$ line profiles of model 5  for different inclination angles $i$ between the polar disk axis and the viewing
direction (10, 20, 40, and 60$\degr$) are shown in Fig.~\ref{BRG}. They suggest that a small inclination of the circumstellar disk ($20 \pm 10
\degr$) is needed to explain the observed line profiles (Fig. 1).

In our model calculations, we calculated more than 100 different models to examine the parameter space described in Table~\ref{tab3}. Five examples
of these models are listed in Table~\ref{modeltab}. Closest agreement  is obtained for a two-component disk model (with two temperature-gradient
regimes) consisting of two zones from $R_{in}$\,=\,0.3\,AU to $R_S$\,=\,0.9\,AU and from $R_S$\,=\,0.9\,AU to $R_{out}$\,=\,3\,AU separated at
$R_{s}$\,=\,0.9\,AU with inner and outer radii $R_{in}$\,=\,0.3\,AU
and $R_{out}$\,=\,3\,AU, respectively, and a power-law temperature distribution in each zone given by \\


Zone 1: \hspace{0mm} $T_d$ = $T_{in}(R/R_{in})^{\alpha_1}$; $T_{in}$ = 1800~K; $\alpha_1$ = $-0.5$; \\

Zone 2: \hspace{0mm} $T_d$ =$T_{in}(R_{s}/R_{in})^{\alpha_1} (R/R_{s})^{\alpha_2}$; $\alpha_2$ = $-0.33.$       \\

\noindent If we  choose, for example, a stronger temperature gradient, then the effective average disk size becomes smaller leading to model
visibilities that are higher than those observed. The calculations show that a similar agreement between model and observations can also be obtained
if the disk parameters vary within the  ranges given by $R_{in}$\,=\,$0.4 \pm 0.1$\,AU, T$_d(R_{in})$\,=\,$1700 \pm 100$\,K, and $\alpha_1$\,=\,$0.5
\pm 0.05$ up to $R$\,=\,0.9\,AU and $\alpha_2$\,=\,$0.4 \pm 0.05$ for $R \ge$ 1\,AU.

\begin{table}[h]
  \centering
  \caption{\label{tab3} Range of parameter variations for our continuum-disk plus disk-wind model calculations}

\begin{tabular}{c|c|c}
\hline \hline
Parameters & Range& Model 5  \\
 \hline
 Disk: &&\\
  $R_{in}$ & 0.25--3 AU (8.8--105~$R_*$)& 0.3 AU (10.5 $R_*$) \\
  $R_{out}$ & 1--5 AU (35--175~$R_*$)& 3 AU (105 $R_*$) \\
  $R_s$ & 0.85--1.25 AU (30--44~$R_*$)& 0.9 AU (31.5 $R_*$)  \\
  $  \alpha_1$ & $-0.4$--~$-0.75$& $-0.5$ \\
  $  \alpha_2$ & $-0.34$-- $-0.4$& $-0.33$ \\
  $T_{in}$ & 1400--2000 K& 1800 K\\
&\\
Disk wind: &&\\
   $\omega_1$ & 0.1--3~AU (3.5--105~$R_*$)& 0.5 AU (17.5 $R_*$) \\
   $\omega_N$ & 0.5--5.7 AU (17.5--200~$R_*$)& 1 AU (35 $R_*$) \\
   $  \gamma$ & $-1$--\,$5$& 2 \\
  $f $ & 0.5--3& 0.5--3 \\
  $  \beta$ & 0.3--2& 1 \\
  $ \theta_1$ & 10$\degr$--80$\degr$& 80$\degr$ \\
  $  \dot{M}_w$ & $10^{-9}$--$10^{-6}M_\odot$yr$^{-1}$   & 10$^{-7}M_\odot$\,yr$^{-1}$
 \\
\hline
\end{tabular}
\end{table}

The calculations of the intensity distributions of the Br$\gamma$-emitting   region and corresponding interferometric model quantities (i.e.,
visibilities, wavelength-differential phases, and closure phases of the model images) show that they are very sensitive to the geometrical and
kinematical parameters of the disk-wind model. Furthermore, they also depend on the disk inclination. We used the following approach to find models
that can approximately reproduce the observations:

1) To select the most appropriate the disk parameters, we calculate the continuum intensity of the disk, $I_d$, at wavelengths near Br$\gamma$ and
compare it with the stellar intensity $I_*$. A good solution has to reproduce the observed ratio $I_d/I_*$, the observed $K$--band flux, and the
observed visibilities.

2) We determine the velocity and density distribution along each  streamline corresponding to the wind geometry and kinematics parameters (see
Sect.~\ref{diskwind} and Appendix A).

3) Using the velocity and density distribution, and choosing the temperature law  along each streamline (we use a constant electron temperature of
8000~K), we solve the equations of the statistical equilibrium for the hydrogen atoms and compute the population of the hydrogen atomic levels and
the ionization degree along each streamline.

4) We then calculate the Br$\gamma$ line profile and compare it with the observed line profile. The solution is good if the theoretical line profile
is in good agreement with the intensity and shape of the observed one.

5) For each disk-continuum and disk-wind model (see Table~\ref{tab3}), we compute the two-dimensional intensity distribution map for several
inclination angles. We calculate all model intensity distributions for 15 different wavelengths across the Br$\gamma$ line and the continuum. The
spectral resolution of AMBER of $R$ = 12\,000 is modeled in the final model intensity distributions (see discussion on the importance of modeling
AMBER's spectral resolution by convolution in Section \ref{ip} and Fig.~\ref{mod1b}).

6) Finally, we compute visibilities, wavelength-differential phases, and closure phases of the obtained 2-D model intensity distributions, which are
needed for the comparison with the observations.


\subsection{Comparison of the models with the observations }\label{compar}

\begin{figure*}
\centering

\vspace*{-8mm}
\hspace*{-8mm}
\includegraphics[width= 198mm, angle =0]{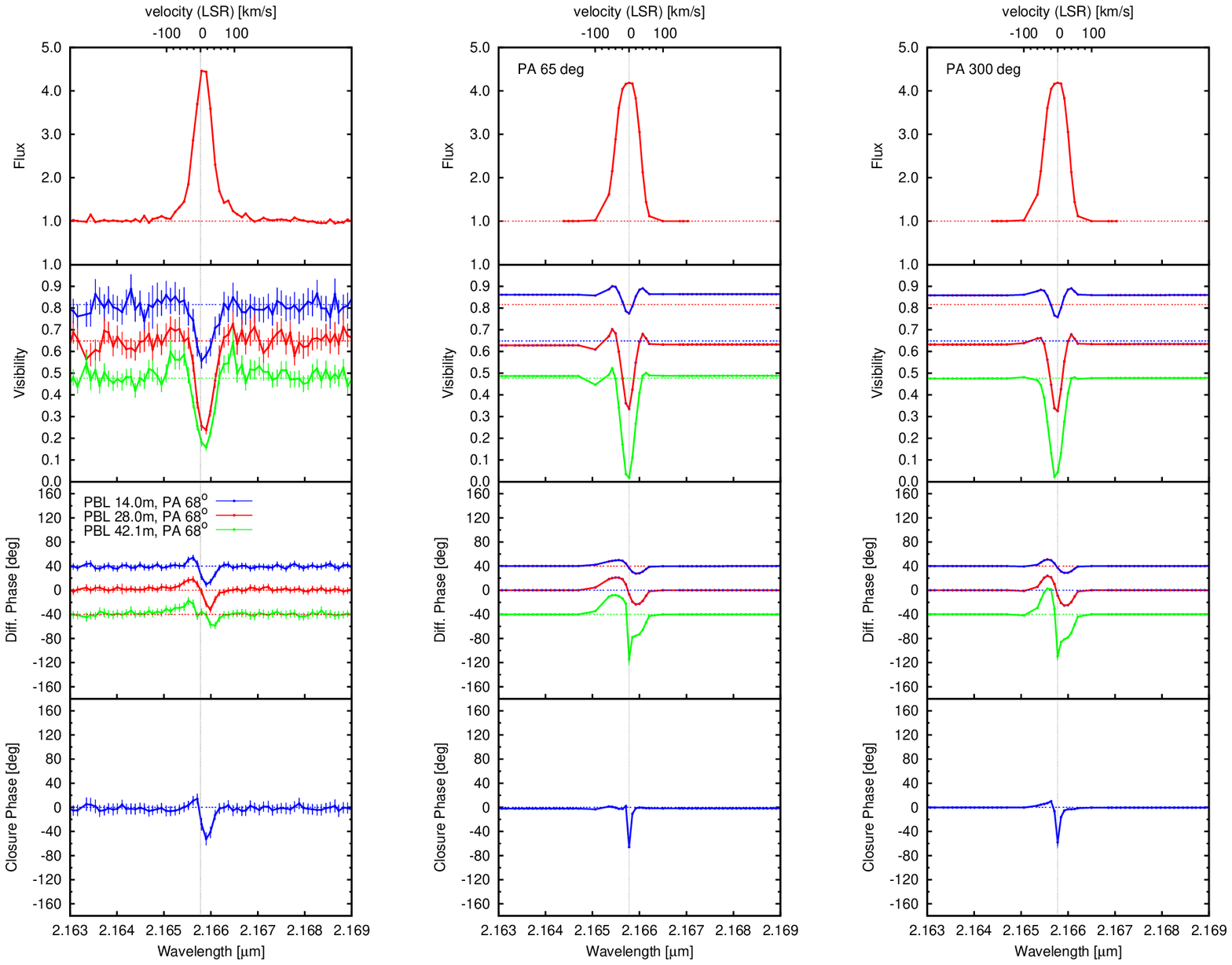}

\vspace*{-122mm}
\caption{\label{mod1} Comparison of the observations (left panel; see Fig.~\ref{obs_fig1}) with the corresponding model quantities of model 5 for
two different PAs of the model image (middle and right panel; the model quantities are calculated for AMBER's spectral resolution of  12\,000, as
discussed in Sect.~\ref{ip} and Fig.~\ref{mod1b}). The middle and right panels show the dependence of the interferometric observables (spectrum,
visibilities, wavelength-differential phases, and closure phases) of our best-fit disk-wind model 5 (disk-wind emitting region, continuum accretion
disk, plus central star; Tables~\ref{tab3} and \ref{modeltab}) on the wavelength across the Br$\gamma$ line for an inclination angle of $i$ =
20$\degr$, clock-wise motion of the disk wind, and for two different PAs of the projected disk polar axis on the sky: 65$\degr$ (middle)  and
300$\degr$ (right).  Several other PAs that are also approximately in agreement with the observations are discussed in the text and  Figs.~\ref{A8}
and \ref{A8a}. The detailed dependence of the interferometric observables on the PA is presented in Figs.~\ref{A8} and \ref{A8a}. }
\end{figure*}

Using this method, we found that model 5 (see Tables~\ref{modeltab} and \ref{tab3}) provides the closest agreement with the AMBER data (see
Figs.~\ref{mod1}, \ref{A8}, and \ref{A8a}) and  the $K$-band continuum flux, as discussed below.  The inner radius of the disk-wind ejection region
of model 5 is 0.5~AU ($\omega_1$ = 17.5~$r_*$) and the half-opening angle of the disk wind is 80$\degr$.  The disk-wind half-opening angle is defined
as the angle between the rotation axis and the innermost wind streamline (see Fig.~\ref{sketch}).  Figure~\ref{mod1} shows flux, visibilities,
wavelength-differential phases, and closure phases of model 5 for the same interferometric baselines as the AMBER observations presented  in
Fig.~\ref{obs_fig1}.

The spectral energy distribution of model 5 predicts that $\log{\lambda F_\lambda}$[erg cm$^{-2}$ s$^{-1}]$ = $-7.13$ in the $K$-band continuum near
Br$\gamma$ for a distance of 250 pc, which is in good agreement with the dereddened SED of MWC~297 from \citet{ack08} and \citet{Alonso09}. The model
ratio $I_d/I_*$\,=\,10 also agrees with the observations.


\begin{figure*}
\centering

\vspace*{-8mm}
\hspace*{-3mm}
\includegraphics[width= 188mm, angle =0]{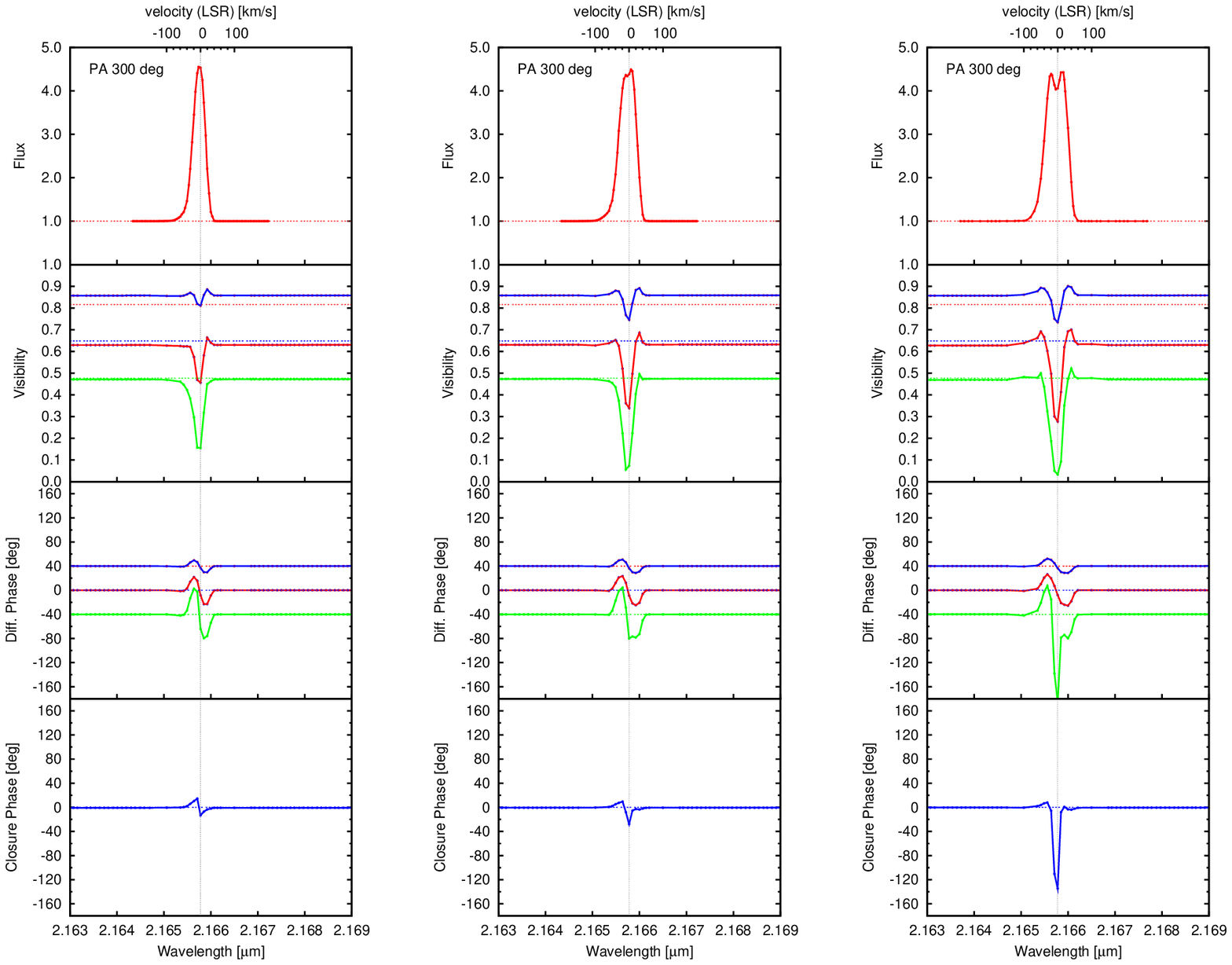}

\vspace*{-116mm}
\caption{\label{mod1a} Dependence of the interferometric quantities (visibilities, wavelength-differencial phases, and closure phases) of our
best-fit disk-wind model 5 (see Table \ref{modeltab}) on the inclination angle $i$: 10, 15, and 20$\degr$ from left to right for the PA 300$\degr$ of
the disk (see Fig.~\ref{mod1}). In contrast to Fig.~\ref{mod1}, the Br$\gamma$ line profiles are given without convolution with the instrumental
profile of the AMBER spectrograph for clearer visibility of the theoretical line profiles. }
\end{figure*}

Since the position angle of MWC~297's projected polar disk axis on the sky is unknown, the interferometric observables of the best-fit model 5
(spectrum, visibilities, wavelength-differential phases, and closure phases) were calculated for all PAs in steps of 10$\degr$  (see Sect.~\ref{ip}).
Figure~\ref{mod1}  compares  the observations (left panel) with two of the models that are approximately in agreement with all observables (clockwise
motion, $i$ = 20$\degr$, middle: PA = 65$\degr$, right: PA = 300$\degr$). A similar agreement can also be obtained for some smaller inclination
angles (see Fig.~\ref{mod1a}).

The detailed PA dependence is presented in Sect.~\ref{ip} and  Figs.~\ref{A8} and \ref{A8a}, which show the following results. For  models with
clockwise motion of the disk wind and $i\sim 20\degr$ (i.e., almost pole-on), we obtained the closest agreement between model and observations for
the PAs of approximately 65, 120, 245, and 300$\degr$ ($\pm 5 \degr$; see Fig.~\ref{A8}). For the models with anti-clockwise motion and $i\sim
20\degr$, we obtained the best agreement for the PAs of approximately 15, 70, 195, and 250$\degr$ (see Fig.~\ref{A8a}).

\section{Discussion}
Our disk-wind model allows us to approximately reproduce all interferometric observables, including the remarkably strong differential and closure
phases measured in our AMBER/VLTI observations of MWC~297 and to draw several conclusions:

1) The AMBER {\it continuum visibilities} confirm previous results (\citealt{MONNIER05,mal07,ack08,kra08a}) that the continuum-emitting region is
remarkably compact. We used geometric models consisting of a stellar point source, an extended halo, and a dominant compact Gaussian or ring to
characterize its size. For the compact ring component, we obtained an inner ring-fit radius of $\sim$2.2~mas ($\sim$0.56~AU). This compact ring
radius is about 5.4~times smaller than the 3~AU dust sublimation radius expected for silicate grains, if there is no radiation-shielding material
between the star and the dust rim.

2) The Br$\gamma$ {\it emission line region} is far more extended than the compact continuum-emitting region. The strong and wavelength-dependent
closure phase reveals that the line-emitting region is very asymmetric along our measured interferometric position angle and the asymmetry
dramatically changes across the Br$\gamma$ line. At the center of the Br$\gamma$ line, we derived a Gaussian fit HWHM radius of $\sim$6.8~mas
(1.6~AU), which is $\sim$2.7 times larger than the compact 0.56~AU continuum Gaussian.  The diameters of the Br$\gamma$-line-emitting region shown in
Fig.~\ref{brg} are derived from continuum-compensated visibilities. Therefore, they are the diameters of the Br$\gamma$-line-emitting region (not of
the emission region of both line and continuum).

3) To interpret our AMBER observations, we employed a magneto-centrifugally driven disk-wind model. This model consists of an accretion disk, which
emits the observed continuum radiation, and a disk wind, which emits the Br$\gamma$ line. The disk wind starts at a distance of $\sim$0.5~AU from the
star, which is further out than the typical value of 0.07~AU suggested for the inner radius of T Tauri star disk winds  \citep{Saf93a,fer07}. The
required large half-opening angle ($\sim$80$\degr$) of the wind in our model is much larger than the typical value (30--45$\degr$) considered for T
Tauri stars (see, e.g., \citealt{Saf93a,fer07}). Both of these features of the model are likely to be caused by the strong influence of the radiation
pressure of the star \citep{DREW98}.

The model images of the disk wind are highly asymmetric and their photocenter shift and shape strongly change across the line profile, resulting in a
strong wavelength dependence of the visibilities, wavelength-differential phases, and closure phase. This asymmetry of the model images is a result
of the Doppler shift of the radiation caused by  the gas motion in the emitting region. The Doppler shift strongly depends on the position within the
disk and  the disk inclination angle.

The properties of the inner region of the circumstellar environment of MWC~297 are remarkable in many respects. In the NIR, it radiates at
temperatures of $T_d \le 2000$~K at a distance of only $\ge$0.5 AU from the star. If the disk is optically thick between the star and the sublimation
zone (i.e., the disk is able to shield the radiation from the cental star), the radius of this zone, $R_{\rm sub}$, will be smaller than for the
optically thin limit. Using the model of optically thick and geometrically thin disks \citep{AS86} heated by the central star, one obtains $R_{\rm
sub}$\,$\sim$\,0.8~AU for MWC~297, which is comparable to the inner disk radius of the best-fit model 5.

The mass-loss rate of the disk wind of the best-fit model 5 is 10$^{-7} M_\odot {\rm yr}^{-1}$ (Table~\ref{modeltab}). The theory of
magneto-centrifugal disk winds predicts a typical ratio $\dot M_w/\dot M_{acc} $ on the order of 0.1 \citep{kon00}. In this case, the mass accretion
rate in the disk of MWC~297, $\dot M_{acc}$, is $\sim$10$^{-6} M_\odot {\rm yr}^{-1}$. According to \citet{MUZEROLLE04}, such a disk is optically
thick between the star and the sublimation zone because of the gas opacity.

What type of matter radiates in this compact inner disk inside the dust sublimation radius? We assume that it is a mixture of the warm, mostly
neutral molecular gas plus refractory (e.g. graphite) grains. Similar properties of the inner compact material were found by \citet{ben10} for the
HAeBe star HD~163296. Interaction of the dust grains with the stellar radiation possibly plays an important role in the acceleration of dust and gas
and the formation of the disk wind.

As shown above, an inner gap in the disk of MWC~297 is needed to explain the observations.  This gap region may be filled with material that is
either fully transparent or semi-transparent in the infrared (e.g., \citealt{tan08}). The formation of this gap could be a result of binarity
\citep{AL94}. Another possibility for gap formation is the interaction of stellar radiation and wind with the inner disk. The strong radiation
pressure and the stellar wind can blow away the disk atmosphere   \citep{DREW98}. As a result, the disk can partially or fully dissipate in the
vicinity of the star.

4) The critical rotation velocity $v_{crit} = \sqrt(2GM_*/(3R_*)$ \citep{mae00} of MWC~297  would be 450~$\kms$ for $R_*$ = 6~$R_{\odot}$ and $M_*$ =
10~$M_{\odot}$, as adopted in our paper. For our determined inclination angle $i = 20 \pm 10 \degr$, the maximum value of $v_{rot}\,\sin{i}$
 would be $150 \pm 80~\kms$  for $v_{rot} <  v_{crit}$. This is in clear contradiction with the value of
$v_{rot}\,\sin{i} = 350 \pm 50~\kms$   obtained by \citet{dre97} for MWC~297.

What is the cause of this discrepancy? Is it our poor knowledge of either mass or radius, the inclination angle of $20 \pm 10 \degr$, or the observed
$v_{rot}\,\sin{i} = 350 \pm 50~\kms$?  The value of $i$ is constrained not only  by  our modeling but also by the recent interferometric studies
reported by \citet{mal07} and \citet{ack08}.  Nevertheless, further studies of $i$ will be required to  constrain $i$ more reliably. Furthermore, it
cannot be excluded that the value of $v_{rot}\,\sin{i} = 350 \pm 50~\kms$  is overestimated, since it is difficult to derive the exact rotational
velocity from the width of the observed absorption lines because of the medium spectral resolution employed and the quite low SNR of the Drew et al.
spectrum caused by the relatively faint magnitude of $V$ = 12.2. Moreover, the absorption lines in MWC 297 might be: a) contaminated by emission from
outflowing material, as is often the case in massive young stars, or b)  formed at a certain distance from the photosphere, where the broadening of
the spectral line would be dominated by the outflow kinematics. Nevertheless,  this question is far from being solved, and we encourage independent
determinations of $v_{rot}\,\sin{i}$ using, for instance, high-resolution spectroscopy and state-of-the-art atmospheric modeling. Future
multi-wavelength observations of MWC~297 using high spectral and spatial resolutions, combined with non-LTE radiative transfer modeling of the
stellar atmosphere, will be essential to improve our knowledge of the stellar parameters such as the rotational velocity.

5) According to our data, the Br$\gamma$ line has a low redshift in the LSR frame (but not in the heliocentric system), in contrast to the model
spectrum. However, it is not yet known whether the Br$\gamma$- emitting region is really redshifted with respect to the star, since the small
observed redshift could  simply be caused by the (unknown) radial velocity of the star  (see  Appendix D).

The model of the emitting region of MWC~297 proposed in this paper needs to be improved. In particular, we need to investigate the dynamics of the
magneto-centrifugal disk wind in the presence of the strong radiation field of the star. It is also necessary to investigate the effect on the
thermal regime of the wind of both  stellar radiation and ambipolar diffusion.

Additional information about the properties of the emitting region can be obtained by analyzing other hydrogen lines, first of all the H$\alpha$
line. The observations by Drew et al. (1997) show that in the spectrum of MWC~297, this line is broader than the Br$\alpha$ line and has symmetric,
extended wings up to 400--500~$\kms$. The Br$\gamma$ line is also much narrower than H$\alpha$. Our preliminary test calculations show that this
difference can be the result of opacity effects ($\tau$\,(H$\alpha$) $\gg \tau$\,(Br$\gamma$)) in addition to Stark broadening. A more detailed
discussion of the H$\alpha$ line properties as well as those of other Balmer lines will be presented in a forthcoming paper.

\section{Conclusions}\label{con}
We have presented VLTI/AMBER observations of the Herbig Be star MWC~297 with a spectral resolution of 12\,000. This high spectral resolution has
allowed us to study the structure and kinematics of the inner sub-AU accretion-ejection region in approximately $\sim$10 different spectral channels
(radial velocities) across the resolved Br$\gamma$ emission line and  compare the observations with disk-wind models.

The $K$-band continuum observations confirm that MWC~297's continuum emitting region is remarkably compact. We measured a ring-fit radius of
$\sim$2.2 mas ($\sim$0.56~AU) for the dominant compact component. At all Doppler velocities across the Br$\gamma$ line, the line-emitting region is
both asymmetric (as indicated by the strong closure phase) and more extended than the compact continuum-emitting region. At the center of the
Br$\gamma$ line, we derived a Gaussian continuum-compensated radius (HWHM) of 6.3\,$\pm$\,0.4~mas (1.6\,$\pm$\,0.1~AU) for the line-emitting region,
which is $\sim$3 times larger than the 0.58~AU HWHM of the compact continuum-emitting Gaussian component.

To interpret our AMBER observations, we employed a magneto-centrifugally driven disk-wind model consisting of an accretion disk, which emits the
observed continuum radiation, and a disk wind, which emits the Br$\gamma$ line. We have demonstrated that all observables, i.e., the observed
Br$\gamma$ line profile, $K$-band continuum flux, visibilities, wavelength-differential phases, and remarkably strong closure phases, can
simultaneously be reproduced by our disk-wind model. The modeling also shows that disk-wind models developed for T Tauri stars can  be used to
interpret Herbig Be star observations if suitable modifications are applied. The modification has to take into account the higher radiation pressure
from hot stars, which can  change the inclination angle of the streamlines  of the disk wind and cause a flatter disk wind  than in T Tauri stars.
The disk wind of MWC~297 starts at a distance of $\sim$0.5~AU ($\sim$17.5 stellar radii) from the star, which is a larger distance than the typical
value of 0.07~AU predicted for T~Tauri stars.


\begin{acknowledgements}

We thank the ESO VLTI team on Paranal for the excellent collaboration. The data presented here were reduced using the publicly available
data-reduction software package {\it amdlib} kindly provided by the Jean-Marie Mariotti Center (http://www.jmmc.fr/data\_processing\_amber.htm). The
telluric spectra used in this work for spectral calibration of the AMBER data were created from data that was kindly made available by the NSO/Kitt
Peak Observatory. A.M., V.G., and L.T. thank the Max-Planck-Society for support during their stay in Bonn. This publication makes use of the SIMBAD
database operated at CDS, Strasbourg, France. We are grateful to Rene Oudmaijer and Bram Acke for sharing their spectroscopic data. Finally, we thank
the referee for his comments that helped to improve the manuscript.
\end{acknowledgements}

\appendix

\section{Description of the model calculations}
\subsection{Disk wind}
\begin{figure*}[t]
\begin{centering}\label{sketch}
\vspace{8mm}
\includegraphics[width=41mm, angle =-90]{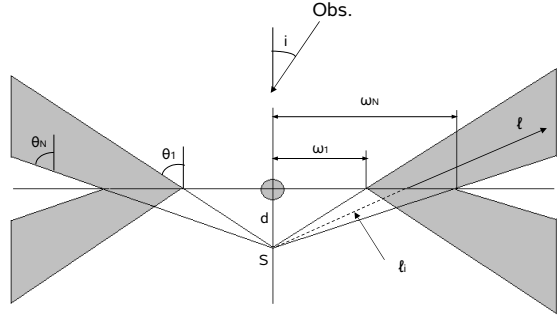}
\vspace*{1mm} \caption{Sketch of the geometry of the disk-wind model adopted.}
\end{centering}
\end{figure*}

For the description of the disk wind, we use a coordinate system ($l$, $\theta$, $\omega$) centered at point S located on the rotation axis (see the
sketch in Fig.~\ref{sketch}). In MHD models of the disk wind, the inclination angle of the first streamline with respect to the disk plane is
typically assumed to be around $60\degr$ \citep{BP82}. However, in the case of Herbig Be stars, the strong stellar radiation pressure possibly bends
down the streamlines and makes the disk wind flatter. Therefore, in our models we use a smaller inclination angle for the innermost streamline (e.g.,
between $10\degr$ and $30\degr$). This effect influences the characteristic size of the wind-launching region.
%


\begin{figure*}[]
\begin{centering}

\hspace*{0mm}
\vspace*{-5mm}
\includegraphics[width= 148mm, angle =-0]{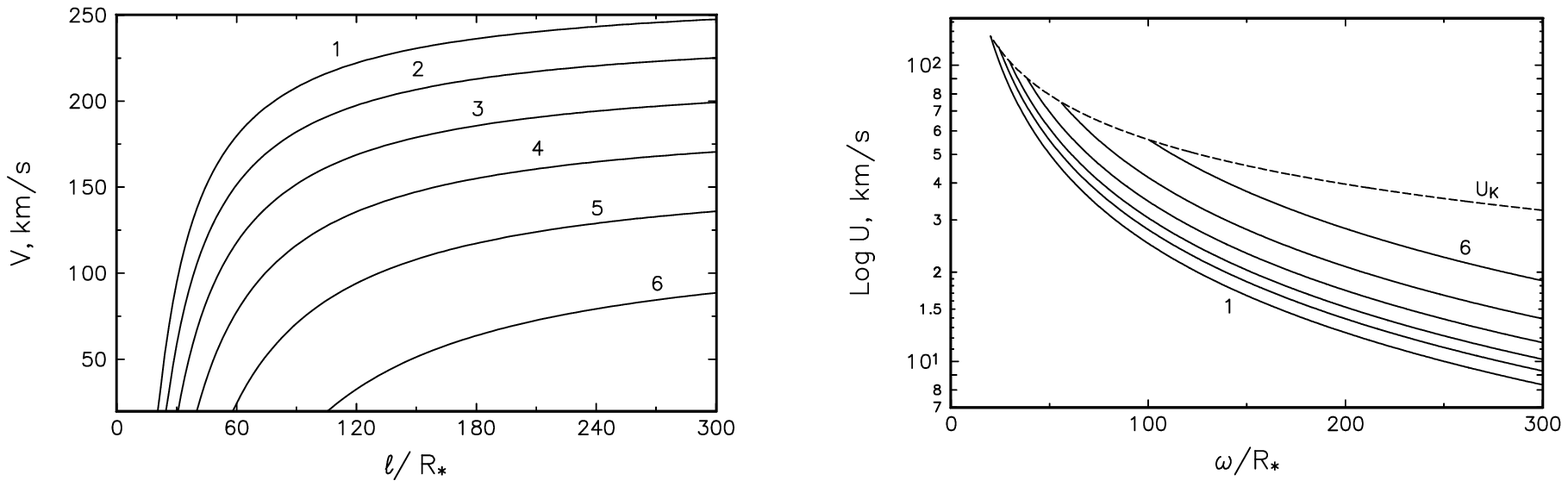}

\vspace*{-157mm}
\caption{\label{vel}{\it Left}: Radial velocity of the outflow along the different streamlines as a function of $l$; numbers 1-6 denote the number
of the streamline. {\it Right}: The same for the rotational velocity; the dashed line shows the Keplerian velocity in the disk.}
\end{centering}
\end{figure*}

As  mentioned in Section 4, the tangential velocity of the wind changes along streamlines as
\begin{equation}\label{velu}
 u(\omega) = u_K(\omega_i)\,(\omega/\omega_i)^{-1}\,,
\end{equation}
where $\omega$ = $l\, \sin{\theta}$ is the distance of the point ($l,\theta$) from the rotation axis,  $\omega_i$ = $l_i\,\sin{\theta}$, and
$u_K\,(\omega_i$) = $(G\,M_*/\omega_i)^{1/2}$ at the point ($\omega_i$) at the base of streamline $i$.

The radial velocity $v$ increases along the streamlines as
\begin{equation}\label{velv}
    v(l) = v_0 + (v_{\infty} - v_0)\,(1 - l_{i}/l)^{\beta}\,,
\end{equation}
where $v_0$ and $v_{\infty}$ are the initial and terminal values of the radial velocity and $\beta$ is a parameter.

We  adopt $v_\infty$ = $f\,u_K(\omega_i)$, where $u_K(\omega_i)$ is the Keplerian velocity at distance $\omega_i$ from the disk axis and $f$ is the
scale factor of the asymptotic terminal velocity to the local Keplerian velocity at the wind emerging point   (see \citealt{KHS06}). The parameter
$f$ can change with distance $\omega$ because the magnetic field in the accretion disk decreases with $\omega$. An example of the velocity field of
one of the models considered below is shown in Fig.~\ref{vel}.

The kinematical model described above is typical of disk-wind models adopted for different astrophysical objects with accretion disks including T
Tauri stars \citep{KHS06}, cataclysmic variables \citep{SV93}, and even active galactic nuclei \citep{MC97}. In these models, the gas streams that
start in the innermost regions of the disk-wind launching region have the highest radial velocities. These kinematical models are characterized by
non-local radiative coupling between distant parts of the moving media \citep[see, e.g.,][]{GG75,RH78}, which complicate the radiative transfer
problem. In our calculations, we neglect the influence of the non-local radiative coupling since its effect on the radiative excitation in the
flattened emitting regions (such as accretion disks or disk winds with large opening angles) is small.

For the calculation of the density distribution $\rho$ in the disk wind, we use the mass-loss rate per unit solid angle, $\dot M_w(\theta)/4\pi$. One
can show that $\dot M_w(\theta)$ and the mass-loss rate per unit area of the disk, $\dot m_w$, are related by
\begin{equation} \dot M_w(\theta) = 4\pi d^2\dot m_w(\omega)\,\cos^{-3}{\theta}\,,
\end{equation}
where $d$ is the distance between point $S$ and the star. The total mass-loss rate is then
\begin{equation}
\dot M_w = \,\int_{\theta_1}^{\theta_N} \dot M_w(\theta)\,\sin{\theta}d\theta.
\end{equation}
In this case, we can write the continuity equation in its usual form
\begin{equation}
4\pi\rho(l,\theta) v(l,\theta) l^2 = \dot M_w(\theta).
\end{equation}
Using this equation, one can calculate the distribution of the number density at each point in the disk wind.

\begin{figure}
\centering \vspace*{-7mm}

\hspace*{-5mm}
\includegraphics[width= 100mm, angle =0]{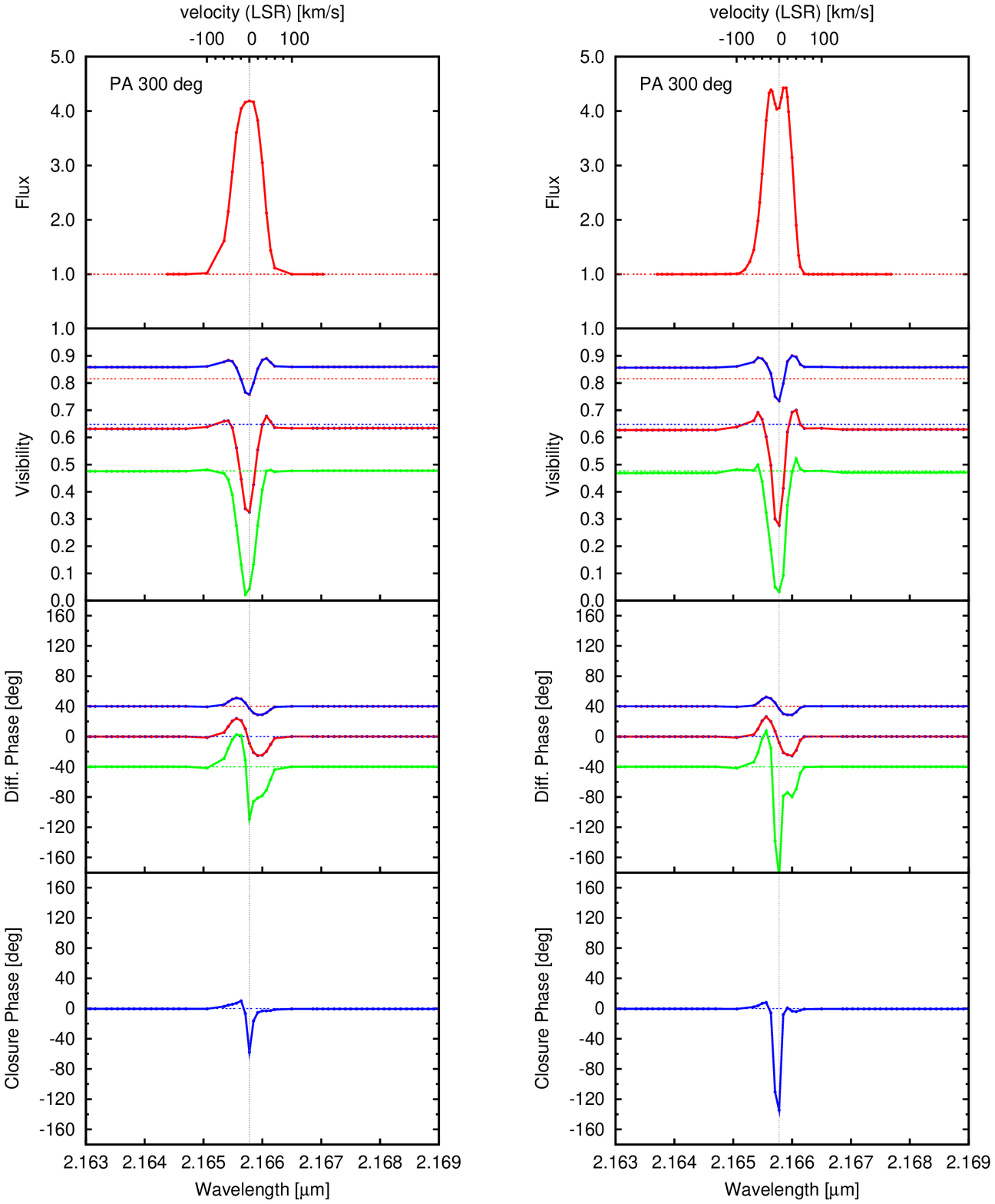}

\vspace*{-22mm} \caption{\label{mod1b} Comparison of the theoretical interferometric quantities of model 5 obtained with (left) and without (right)
modeling  of AMBER's spectral resolution of 12\,000. In both cases, position and inclination angles are identical (PA = 300$\degr$, $i$ =
20$\degr$).}
\end{figure}
%

\subsection{Calculation of the model intensity distributions}
For the calculations of the 2-D intensity distributions (maps) of the emitting region, we use the coordinate system ($x,y,z$) centered on the star.
The ($x,y$) plane coincides with the sky plane, the $x$ axis is the intersection of the disk plane with the sky plane, and the $z$ axis is parallel
to the line-of-sight. In this case, the intensity of radiation at a frequency $\nu$ within a spectral line is
\begin{equation}\label{eqI}
I_{w}(\nu,x,y) =\int_{z_{min}}^{z_{max}}S(\textsf{r})\phi(\nu-\nu_0\frac{\textsf{v}_z(\textsf{r})}{c})\,
e^{-\tau(\nu,\textsf{r})}\kappa(\textsf{r})dz,
\end{equation}
where $\textsf{r}$ is a vector, $|\textsf{r}|$ = $(x^2 + y^2 + z^2)^{1/2}$, $\textsf{v}_z(\textsf{r})$ is the projection of the velocity at point
$\textsf{r}$ on the line-of-sight, and $S$ is a source function for a transition between energy levels $i$ and $j$
\begin{equation}\label{eqS}
S(\textsf{r}) =\frac{2h\nu^3}{c^2}\,\left(\frac{n_j(\textsf{r})} {n_i(\textsf{r})}\frac{g_i}{g_j}-1\right)^{-1}\,,
\end{equation}
$\kappa$ is the integrated line opacity in the considered spectral line, $\tau(\nu,\textsf{r})$ is the line optical depth at point $\textsf{r}$ and
for the frequency $\nu$ in the direction to an observer, and $n_i$ is the number density of atoms in the $i$-th  state
\begin{equation}\label{tau}
\tau(\nu,\textsf{r}) = \int_{z}^{z_{max}}\kappa(\textsf{r}')\phi(\nu-\nu_0\frac {\textsf{v}_z(\textsf{r}')} {c})dz'\,,
\end{equation}
where $\phi$ is the profile function normalized to unity,  $\textsf{r}'$ is the vector with coordinates $(x,y,z^\prime)$, and $z^\prime$ changes from
the current value $z$ in Eq.~\ref{eqI} to the outer boundary value $z_{max}$.

The optical depth of the disk wind in the line frequencies in the direction toward the star is $\tau_*(\nu)$ = $\tau(\nu,\textsf{r} = 0)$. We assume
that there is a complete redistribution of the line frequencies in the reference frame of the atom and use the Doppler profile $\phi$ in the
calculations of the Br$\gamma$ and the Voigt profile for the $H\alpha$ line (with the same constants as in \citealt{KHS06}).

The calculations of the ionization state and the number densities of the atomic levels were performed in a cylinder with a radius $r_{c}$ up to
300\,$R_*$ and a height $h_{c}$ = 600\,$R_*$ divided into grid cells in $l,\theta$ coordinates. In each cell, we solved the equations of the
statistical equilibrium for the 15-level hydrogen atom + continuum, taking into account both collision and radiative processes of excitation and
ionization (see \citealt{GM90} for more details). We assumed that the distribution of the atomic sub-levels follows their statistical weights.
Johnson collision rates \citep{joh72} were used for all transitions except 1-2; for the latter, we used those from \citet{sch90}.

The main sources of continuum radiation near the Br$\gamma$ line are: i) the circumstellar disk, ii) the stellar atmosphere, and iii) the free-free
and free-bound radiation of the disk wind. If the radiation of the star ($I_*$) intersects the disk wind in the line of sight, it has to be corrected
for the absorption in the wind and we observe $I_*\,exp(-\tau_*(\nu))$. In the adopted coordinate system, we obtain for the observed disk intensity
distribution
\begin{equation}
I_d(\nu,x,y) = B_{\nu}(T_d(R))\,e^{-\tau_{w}(\nu,x,y)},
\end{equation}
where $R$ = $(x^2 + y^2/\cos^2{i})^{1/2}$, $\tau_{w}(\nu,x,y)$ is the optical thickness of the disk wind at frequency $\nu$ between an observer and
the point in the disk with the coordinates $x' = x$ and $y' = y/\cos{i}$, and $i$ is the inclination angle ($i$ = 0 corresponds to the pole-on
orientation). It is assumed that the disk is flat and optically thick; each elementary area of the disk  radiates as a black body with the local
temperature $T_d(\omega)$.

The total intensity map of the object at frequency $\nu$ is the sum
\begin{equation}
      I(\nu,x,y) = I_{w}(\nu,x,y) + I_d(\nu,x,y),
\end{equation}
where $I_{w}(\nu,x,y)$ is the intensity of the disk wind at the point $x,y$ at an emission line frequency $\nu$, and $I_d$ is the disk intensity
distribution.

At the central part of the map where $x^2 + y^2 \le r_*$, we have to add the intensity of the stellar radiation $I_*(\nu) exp(-\tau_*(\nu))$ to this
sum. Our calculations have shown that the contribution of the free-bound and free-free radiation of the disk wind at the wavelength of the Br$\gamma$
line is small compared with the disk radiation.

To model observations with a spectral resolution lower than the theoretical one, the intensity maps have to be convolved with the instrumental
profile

\begin{equation}\label{conv}
      I_c(\nu,x,y) = \int_{-\infty}^{+\infty}I(\nu',x,y) i(\nu'-\nu)\,d\nu',
\end{equation}
where $i(\nu)$ is the normalized instrumental profile of AMBER.

In our case, the spectral resolution is quite high ($\sim$25~$\kms$). The calculations of the theoretical maps (Fig.~\ref{lmap}) and the
interferometric model quantities (Figs.~\ref{mod1}, \ref{A8}, and \ref{A8a}) were made using Eq.~\ref{conv}. A comparison of the results obtained
with and without spectral convolution shows (see section~\ref{ip}) that noticeable differences in the interferometric quantities exist even at this
high spectral resolution.

The flux at distance $D$ from the object is an integral over the entire area in the x,y -plane of the emitting region
\begin{equation}
      F_\nu = \frac{1}{D^2}\int_{A}^{}I(\nu,x,y)dx\,dy.
\end{equation}

To summarize, the model assumptions are:
\begin{itemize}
    \item The disk wind is launched from the accretion disk
    surface in the range from distance $\omega_1$ (the starting point of the
    innermost streamline) to $\omega_N$ (starting point of the outermost
    streamline). The half-opening angle (which is the angle between the
    rotation axis and the innermost streamline) is $\theta_1$ (see Fig. A1).
    \item The wind is assumed to be isothermal with an electron
    temperature of 8000~K (for all models in Table A.1).
    \item The disk wind possesses both radial and tangential (rotation) motion.
    \item The number density distribution is calculated from the
    mass continuity equation for each streamline using the local mass-loss rate
    per unit area.
    \item The atomic level populations were computed in the Sobolev
    approximation; a 15-level hydrogen atom has been considered.
    The 3--D radiative problem (exact integration) has been solved in a
    cylinder with a radius of 300\,$R_*$
    and a height 600\,$R_*$.
    \end{itemize}

%
\subsection{Interferometric model quantities}\label{ip}

\begin{figure*}
\centering \vspace*{-12mm}

\hspace*{-8mm}
\includegraphics[width= 210mm, angle =0]{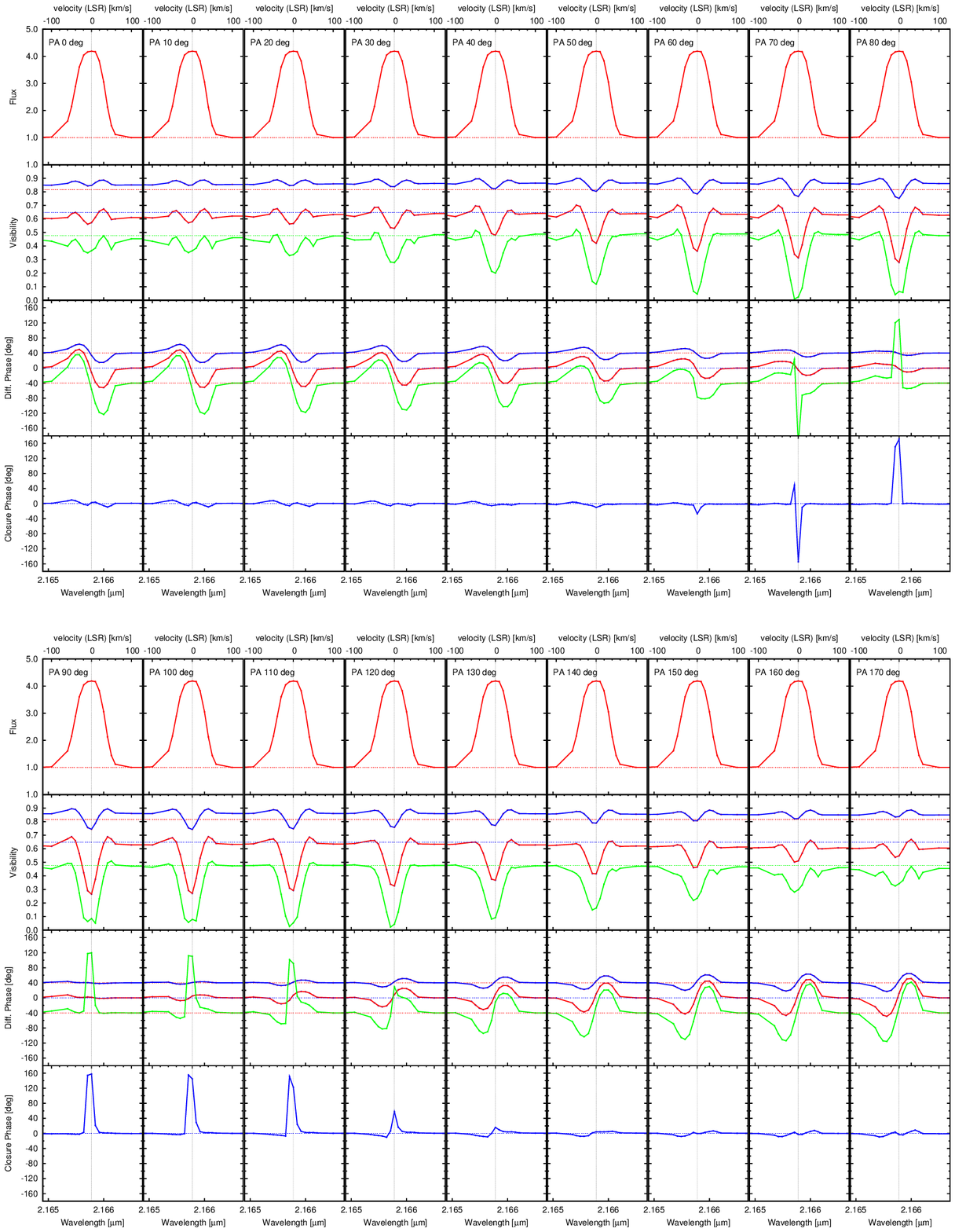}

\vspace*{-50mm} \caption{\label{A8} Dependence of the visibilities and phases of model 5 (see Tables~\ref{tab3} and \ref{modeltab}) on both the
wavelength across the Br$\gamma$ line and  the PA of the model on the sky for an inclination angle of $i$ = 20$\degr$ and clockwise motion of the
disk wind (the corresponding anti-clockwise plots are shown in Fig.~\ref{A8a}). {\it Top}: PAs from 0 to 80$\degr$ in steps of 10$\degr$ from left to
right; {\it Bottom}: PAs = 90 to 170$\degr$. Closest agreement between observation and model is obtained for PAs of 65, 120, 245, and 300$\degr$, as
discussed in Sect.~\ref{ip}.}
\end{figure*}

\begin{figure*}
\centering \vspace*{-12mm}

\hspace*{-8mm}
\includegraphics[width= 210mm, angle =0]{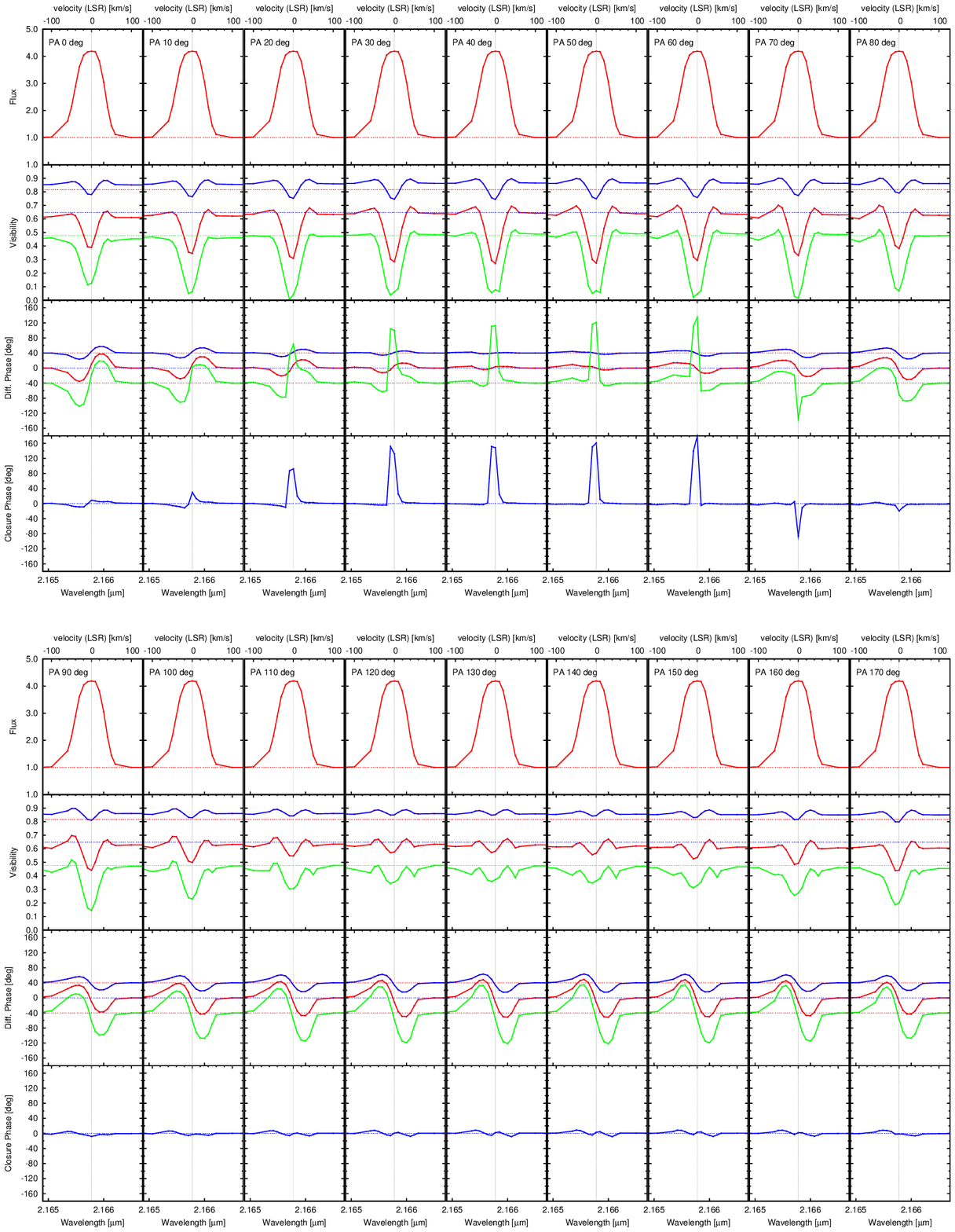}

\vspace*{-50mm} \caption{\label{A8a} Same as Fig.~\ref{A8}, but for anti-clockwise motion. The  best agreement between model and observations is
obtained for the PAs of approximately 15, 70, 195, and 250$\degr$ (see  figure caption \ref{A8}).
 }
\end{figure*}


%
\begin{figure*}
\begin{centering}
\includegraphics[width=108mm,angle=270]{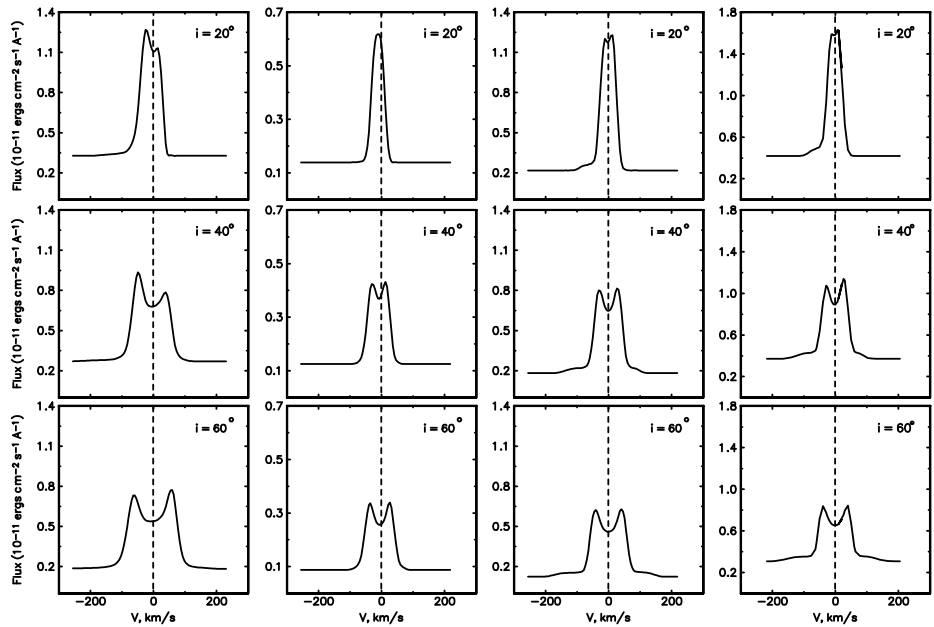}
\vspace{-15mm} \caption{\label{brgmodel} Disk-wind model profiles of the Br$\gamma$ line for different inclination angles (from top to bottom: 20,
40, and 60$\degr$). From left to right, the profiles correspond to models 1, 2, 3, and 4  in Table~\ref{modeltab} (model 5 is discussed in
Fig.~\ref{BRG}).}
\end{centering}
\end{figure*}

\begin{table}[h]
\begin{center}
\caption{Five examples of the computed disk-wind models.}\label{modeltab}
\begin{tabular}{l|c|c|c|r|c|c}
\hline \hline
  Model & $\dot M_w$ & $\theta_{w}$ & $\omega_1$ & $\omega_N$ & $\gamma$ & $f$\\
        & $10^{-8}\,M_\odot/yr$ & deg.     & $r_*$ & $r_*$& & \\
  \hline
  1  & 3  & 70 & 20   & 80 & 0 & 3 \\
  2  & 3  & 10 & 30   & 200& 1 & 2 \\
  3  & 1.2& 80 & 25   & 200& 1 & 2 \\
  4  & 0.6& 80 & 10   & 50 & 0 & * \\
  5  & 10 & 80 & 17.5 & 35 & 2 & * \\
\hline
\end{tabular}
\end{center}
Note: * In models 4 and 5, f is different for different streamlines (see Sect. A.1)
\end{table}

Using the theoretical intensity distribution maps of the emitting regions, one can calculate the interferometric quantities of the model: spectrum,
visibilities, wavelength-differential phases, and closure phases for the same interferometric baseline lengths and position angles as the AMBER
observations. Examples of  interferometric quantities of disk-wind model 5 (see Table~\ref{modeltab}) are shown in Figs.~\ref{mod1}, \ref{mod1a},
\ref{mod1b}, \ref{A8}, and \ref{A8a}.

In Figure~\ref{mod1b}, we show the interferometric model quantities of the Br$\gamma$ line calculated for model 5 with and without convolution of the
theoretical intensity maps with the spectral AMBER profile corresponding to spectral resolution 12\,000 (see Eq.~\ref{conv}). One can see that in
both cases, the visibilities inside the line profiles are very similar for all three baselines.  However, the differential phases for the largest
baseline (42~m) and the closure phases are quite different.

In Figures~\ref{A8} and \ref{A8a}, we show the detailed dependence of the visibilities and phases of model 5 (see Tables~\ref{tab3} and
\ref{modeltab}) on both the wavelength across the Br$\gamma$ line and  the PA of the model on the sky for both an inclination angle of $i$ =
20$\degr$ and clockwise (Fig.~\ref{A8}) and anti-clockwise (Fig.~\ref{A8a}) motion of the disk wind  (the PA is counted from north to east with north
up and east to the left). In Fig.~\ref{A8}, the closest agreement between observation and model is obtained for PAs of 65 and 120$\degr$. The
different sign of the phase at the PA\,=\,120$\degr$ does not disagree with the AMBER observations since the sign is unknown. The plots for the PAs
of 180 to 350$\degr$ are not shown since they give the same results as the PAs from 0 to 170$\degr$, except an opposite sign of the differential and
closure phases, since a 180$\degr$ rotation leads to a conjugate complex Fourier transform. Therefore, the PAs of 65\,+\,180\,=\,245$\degr$ and
120\,+\,180\,=\,300$\degr$ agree with the observations in addition to the PAs of 65 and 120$\degr$. For anti-clockwise motion (Fig.~\ref{A8a}), the
closest agreement between model and observations is obtained for the PAs of approximately 15, 70, 195, and 250$\degr$.

\subsection{ Br$\gamma$ line as a function of the model parameters}
%
Examples of the theoretical Br$\gamma$ line profiles calculated for the models listed in Table~\ref{modeltab} are presented in Fig.~\ref{brgmodel}.
It shows that even for large inclination angles $i$, the profiles are quite narrow. This is due to the large radius of the disk-wind launching
region. Except for a small range of $i$ values close to pole-on view ($i\leq 20\degr$), all orientations of the disk result in a double-peaked
Br$\gamma$ line. In observations, the double-peaked structure can, of course, appear single-lined if the line is narrow and the spectral resolution
is too low.

A single-peaked Br$\gamma$ emission line can also be obtained if a weak, narrow emission component is added to the model. Such  emission can arise at
large distances from the star, as discussed in the following Sect. \ref{lov}.
%

\subsubsection{Low-velocity component of the disk wind }\label{lov}
\begin{figure}[h]
\vspace{15mm}
\begin{centering}
\includegraphics[width= 7.8cm, angle =0]{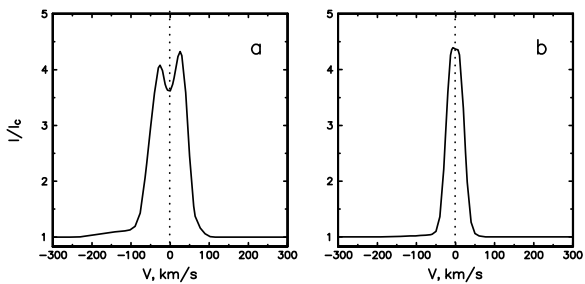}
\caption{\label{inter} Br$\gamma$ line profile of model 4 with  constant (a) and  outward decreasing (b) parameter $f$ ($i$ = 20$\degr$; see
Section~\ref{lov} for details). }
\end{centering}
\end{figure}

A low-velocity outflow from circumstellar disks can be caused by the heating of the upper layers of the disk atmosphere by EUV ($h\nu$ $>$ 13.6 eV)
radiation from  the central star. As a result, the thermal velocity of atoms and ions in the outer part of the disk can exceed the escape velocity.
This is the region where the disk can evaporate (Hollenbach et al. 1994). The inner radius of this zone is
$r_g$ = $GM_*/v_s^2$\,,
where $v_s$ is the sound speed. For ionized gas and $M_*$ = 10~$M_\odot$, $r_g$ is $\sim$7$\times$10$^4/T_e$ $\approx$ 70~AU (\citealt{Gorti09}).
This value exceeds the radius of the AMBER FOV ($\sim$120~mas or $\sim$30~AU at the distance of MWC~297),  meaning that {\it the photo-evaporated
disk wind cannot contribute to the emission observed with VLTI}. However, in the extended region of the disk between the outer radius of the MHD wind
ejection region ($\sim$1~AU) and $r_g$, the EUV photons ionize the upper layers of the disk atmosphere and produce a mini-H II region. This part of
the disk does not evaporate, but  can radiate in the Br$\gamma$ and other lines. This emission is broadened by the Keplerian rotation and thermal
motion of atoms. Its contribution to the total emission may be small, but potentially enough to fill in the small depression in the cental part of
the double-peaked Br$\gamma$ line profile formed by the disk wind. We simulated this profile (Fig.~\ref{inter}) by changing the parameter $f$ from
$f$ = 3 at $\omega_1$ to 0.5 at $\omega_N$. A more detailed modeling of this narrow emission is beyond the scope of this paper and will be presented
in a separate one.

\subsubsection{Screening effects}

\begin{figure}
\centering
\includegraphics[width=0.26\textwidth]{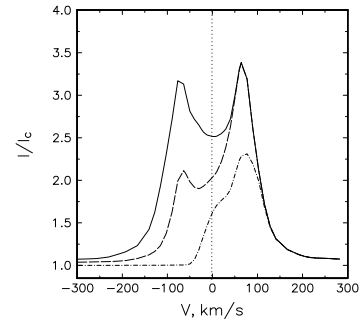}
\caption{\label{screen} Dependence of the Br$\gamma$ line profile on screening caused by an optically thick, dusty disk wind. The solid line
represents the unscreened line profile of model 2 (see Table \ref{modeltab}) with $f$ = 5, while the long-dashed ($H_d$ = 100 $R_*$; $\theta$ =
20$\degr$) and dot-dashed ($H_d$ = 130 $R_*$, $\theta$ = 25$\degr$) lines show models with screening. For all models shown, the inclination angle is
$i$ = 70$\degr$. See text for further details.}
\end{figure}

As noted above, the optically thick part of the disk may screen some parts of the disk wind. This could lead to a red-shift of emission lines
(depending on $i$, see Fig.~\ref{screen}). The calculations show that this effect strongly depends on the opening angle $\theta_1$ of the disk wind.

At large $i$, the approaching part of the disk wind can be screened from an observer by  optically thick layers of the flared disk. We  simulated
this screening effect by introducing a screen with a sharp upper boundary at some distance from the center. The angular size $\theta$ of the screen
seen from the stellar position is the parameter that we introduced. The calculations show (see Fig.~\ref{screen}) that for some $\theta$ (depending
on the disk-wind parameters) and $i \simeq 70\degr$, the screening effect leads to a red-shifted  emission line. It is likely that this mechanism is
not important in our case, since the observations suggest that MWC~297's  inclination angle is small (10 to 20$\degr$).

\section{Wavelength calibration \label{wavecal}}

   \begin{figure}[h]

   \centering
\hspace*{-93mm}
   \includegraphics[angle=-90,width=1.0\textwidth]{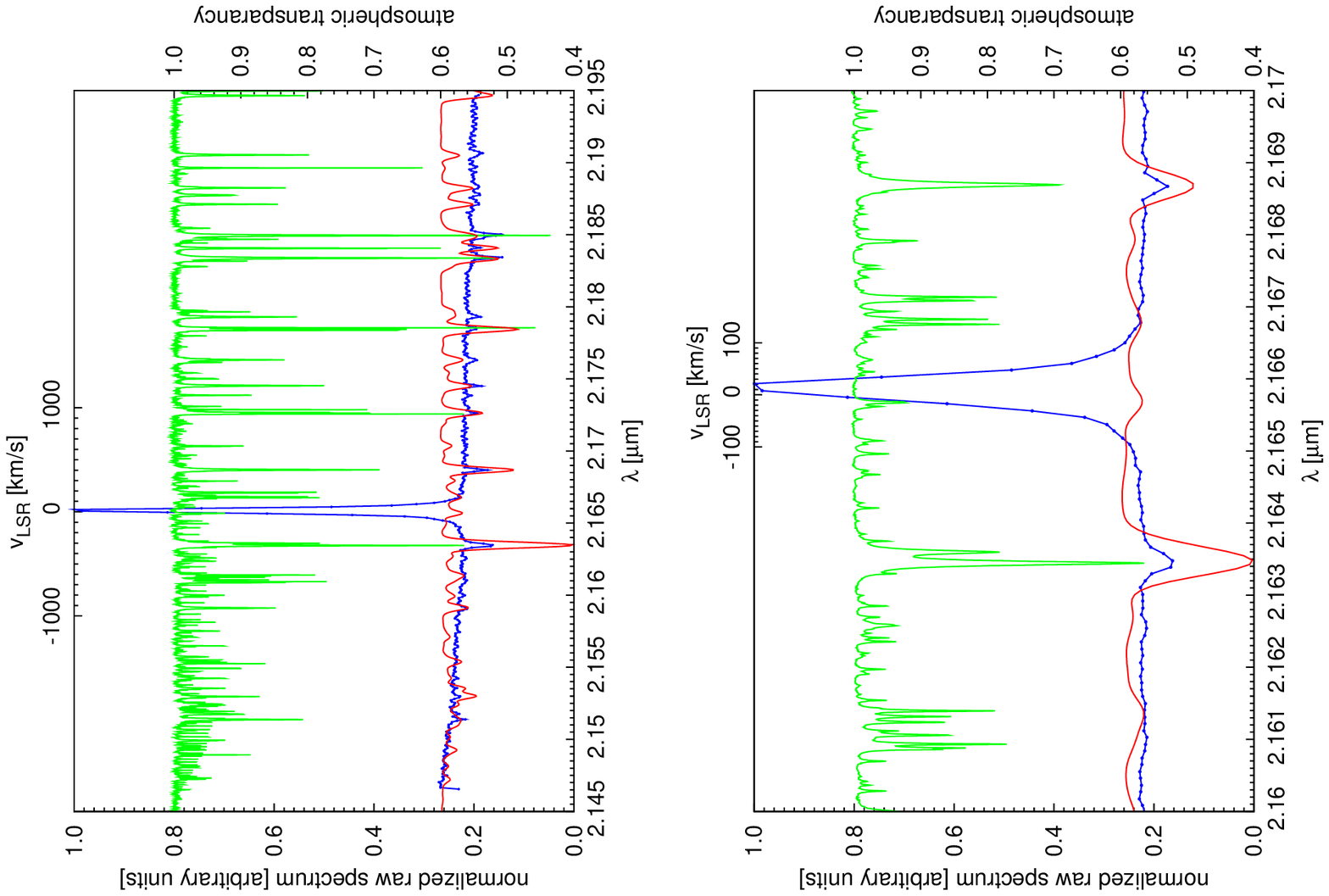}\\

\vspace*{-5mm}
    \caption{\label{groh_wavecal2} Wavelength calibration of the AMBER data.
                {\it Top}: Raw AMBER spectrum of MWC~297 (solid blue line) compared to a telluric
                           spectrum from the Kitt Peak Observatory. The green line shows
                           the original Kitt Peak spectrum with a spectral resolution of 60\,000, while the
                           red line displays the telluric spectrum that has been convolved to the same
                           spectral resolution as the AMBER data ($R=12\,000$). For clarity,
                           the red curve has been shifted down by 0.4. As the figure reveals, after the wavelength
                           calibration, the AMBER raw spectrum of MWC~297 matches several telluric features.
                {\it Bottom}: Same as top panel, but just for the wavelength region around the Br$\gamma$ line.
    }
   \end{figure}

   \begin{figure}[h]
    \vspace*{2mm}
   \centering

   \includegraphics[angle=0,width=0.45\textwidth]{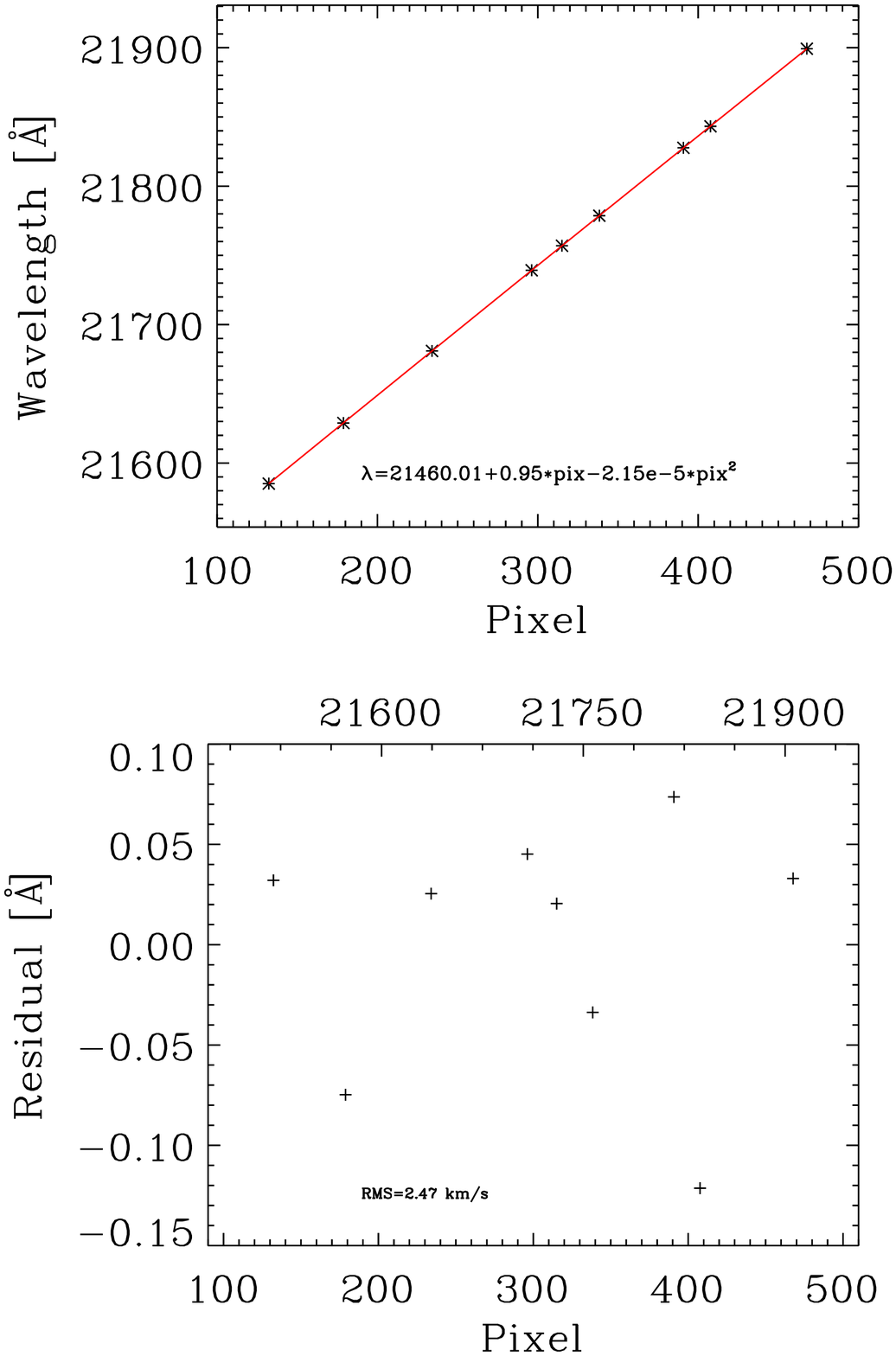}

\vspace*{-4mm}
    \caption{\label{groh_wavecal1}{
\it Top:} Calibration between the pixel and wavelength values of the VLTI/AMBER observations of MWC~297. A least-square polynomial fit of the data is
also shown (red line). {\it Bottom:} Residuals between the measured position of the telluric line spectrum obtained from the wavelength calibration
of the AMBER data and the expected position from the reference telluric spectrum provided by NOAO. The root-mean square value is of the order of
$2.5~\kms$.}
   \end{figure}
Using the telluric spectrum provided by NOAO as a reference (see Fig.~\ref{groh_wavecal2}), standard IRAF routines and specific IDL routines
developed by JHG were used to find a solution to the pixel-to-wavelength calibration of the AMBER spectrum (Fig.~\ref{groh_wavecal1}). The estimated
uncertainty in the wavelength calibration amounts to $\sim$2.5~\kms (Fig. \ref{groh_wavecal2}).


\section{Beam commutation device \label{bcd}}

\begin{figure}[h]
  \centering
  \includegraphics[angle=0,width=0.48\textwidth]{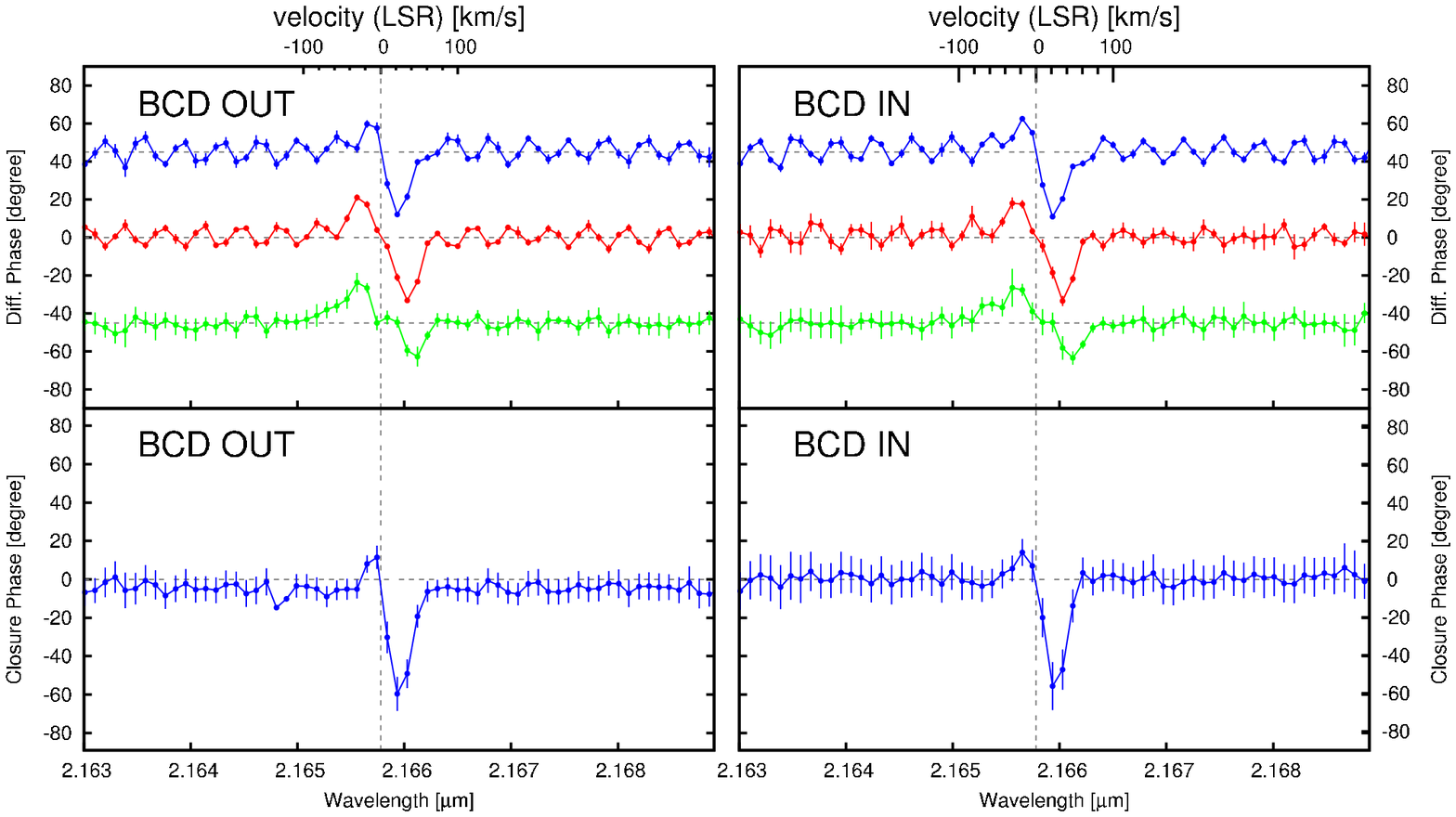}
  \caption{\label{bcdfig}
    Differential phases and closure phases derived from
    the data obtained without ({\it left}, BCD~OUT) and with ({\it right},
    BCD~IN) the beam commutation device.
    The shown signals are not yet corrected using the calibrator data.
    After we corrected the sign changes introduced by the BCD device,
    the two data sets yield the shown signals, which are nearly identical and
    demonstrate the high accuracy of the measured phase signals.
  }
\end{figure}
\begin{figure}[h]
  \centering
  \includegraphics[angle=0,width=0.36\textwidth]{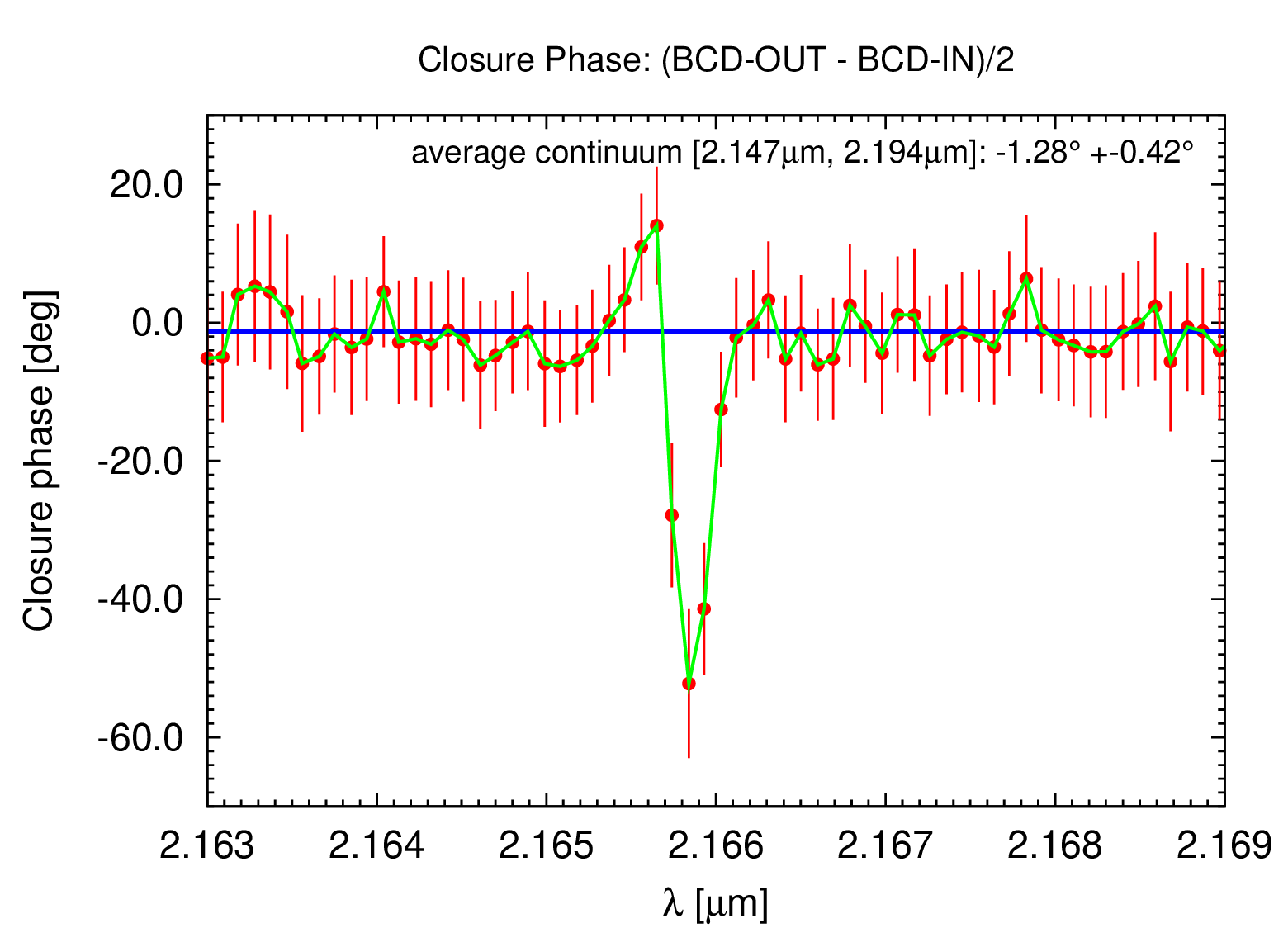}
  \caption{\label{bcdfig2}
    BCD-calibrated closure phase signal obtained by subtracting  the calibrated closure phases obtained from the
    BCD~IN and BCD~OUT data. The blue line is the fit of a constant to the continuum (2.147--2.194$\mu$m)
    closure phase outside the Br$\gamma$ line.
  }
\end{figure}

To optimize the calibration accuracy, we employed the AMBER Beam Commutation Device \citep[BCD,~][]{pet07}, as described in Sect.~\ref{sect_obs}.
This device is located very early in the optical path of the AMBER instrument and allows us to exchange two of the three telescope beams within a few
seconds. Performing this beam commutation  between two subsequent exposures,  allows one to correct potential instrumental drifts. The insertion of
the BCD causes a sign change in one of the three differential phases and   the  CP signal \citep{mil08}. Our BCD~OUT/BCD~IN data sets precisely
showed these expected sign changes. The differential phases and closure phases derived from both independent data sets (BCD~OUT/BCD~IN) are shown in
Fig.~\ref{bcdfig}. Subtraction of the closure phases obtained with and without BCD cancels potential instrumental phase drifts. Fig.~\ref{bcdfig2}
presents the derived closure phase and a fit of a constant to the continuum closure phases. The obtained averaged continuum closure phase is $-1.28
\pm 0.42\degr$.


\section{Optical spectroscopy}\label{optical_spectrum}

MWC\,297 has been rarely observed spectroscopically mostly because of its relative faintness ($V$\,$\sim$\,12.2~mag) and high reddening ($B-V \sim
2.1$~mag). Only two spectra that cover almost the entire optical range, although with a low resolution of 0.8 \AA, have ever been  published
\citep{dre97,andr98}. Higher-resolution spectroscopy was obtained for a few regions, mainly centered on strong emission or absorption lines
(\ion{He}{i} 5876 \AA, \ion{Na}{i} D-lines, [\ion{O}{i}] 6300 \AA, and H$\alpha$, \citealt{zick03,ack05}). A summary of the published optical
spectroscopic data is presented in Table \ref{published_optical_spectra}.

\begin{table}
\caption{Summary of published optical spectroscopic observations of MWC~297} \label{published_optical_spectra} \centering
\begin{tabular}{llrlc}
\hline\noalign{\smallskip}

Date                  & $\lambda\lambda$, &  $R$                     & V                & Ref.\\
                      &  \AA              & $\lambda/\Delta \lambda$ &   km\,s$^{-1}$   &     \\
\noalign{\smallskip}\hline\noalign{\smallskip}

1987 Sept.  9         & 6535--6600        & 23\,000                   &                  & 1   \\
1987 Sept. 13         & 5868--5903        & 45\,000                   & $+6 \pm 1^{\rm c}$ & 1   \\
                      & 6290--6335        &                          &                  & 1   \\
1992 Sept.  9         & 6290--6759        &  8\,000$^{\rm a}$         &                  & 2   \\
1994 June  20--22     & 3876--5198        &  8\,000$^{\rm a}$         & $+10 \pm 8^{\rm d}$& 3   \\
                      & 5840--8809        &                          &                  & 3   \\
1997 July  21, 23     & 4119--5734        &  5\,000$^{\rm b}$         &                  & 4   \\
1997 July  20, 24, 25 & 6297--8911        &                          &                  & 4   \\
2002 Apr.   1         & 6279--6321        &125\,000                   & $+9 \pm 1^{\rm e}$ & 5   \\
2002 June  25         & 5635--6637        & 30\,000                   & $+10 \pm 2^{\rm f}$& 5   \\
\noalign{\smallskip}\hline
\smallskip
\end{tabular}
\begin{list}{}
\item Column 1: observing date; Column 2: observed spectral range; Column 3: average resolving power; Column 4: average LSR radial velocity (see
explanations below and in the text) ; Column 5: reference to the reporting paper. \item $^{\rm a}$ -- two-pixel resolution 0.8 \AA \item $^{\rm b}$
-- 1.3 \AA\ resolution at $\lambda$ = 6000 \AA \item $^{\rm c}$ -- from the [\ion{O}{i}] 6300 \AA\ line \item $^{\rm d}$ -- from emission lines of
\ion{Fe}{ii} 4923 and 5018 \AA\, [\ion{Fe}{ii}] 5159 \AA, \ion{Si}{ii} 5041 and 5056 \AA, and \ion{Si}{ii} 5958 and 5979 \AA \item $^{\rm e}$ -- from
the [\ion{O}{i}] 6300 and 6363 \AA\ lines \item $^{\rm f}$ -- result published by \citet{ack08} \item References: 1 -- \citet{zick03}; 2 --
\citet{vie03}; 3 -- \citet{dre97}; 4 -- \citet{andr98}; 5 -- \citet{ack05}
\end{list}
\end{table}

A detailed study of the object's spectrum is beyond the scope of this paper. Thus, we only focus on the details that are relevant to the modeling of
our interferometric results. One important issue for this purpose is the geometry of the circumstellar environment. The strongest emission-line
profiles exhibit a {\it double-peaked structure}, which can only be revealed at a resolving power of $R \ge 30\,000$ (see \citealt{zick03,ack05}).
This  prompted \citet{zick03} to suggest that MWC~297 might be surrounded by a disk that is viewed at an intermediate inclination with respect to the
line-of-sight.

Another crucial parameter is the {\it systemic (stellar) radial velocity}. Since this had never been carefully discussed, we critically examined
possible methods to derive this quantity and performed an independent study of the published spectra. Absorption lines of the photospheric origin are
only seen in the spectra obtained by \citet{dre97}. They include a few \ion{He}{i} lines blueward of 5800 \AA\ and those of \ion{C}{iii} and
\ion{N}{iii} in the 4500--4700 \AA\ range. All of them are broad, which was interpreted to be caused by a high stellar rotation rate \citep{dre97}.
Additionally, the residual intensities of the carbon and nitrogen lines are within $\sim$5 \% of the continuum, hampering high accuracy measurements
even at the SNR of $\sim$120 achieved in the spectrum. The average LSR radial velocity, which we derived from five \ion{He}{i} absorption lines at
4009, 4026, 4143, 4471, and 4713 \AA, is $-7\pm7$~km\,s$^{-1}$. However, \ion{He}{i} lines in the spectra of B-type emission-line stars are known to
be strongly affected by the circumstellar material in the dense regions close to the star. Even pure absorption lines can exhibit variations that
result in the observed shifts from the photospheric positions (see \citealt{isr96}). Therefore, these lines cannot be considered as a reliable source
of the systemic velocity. The most clearly recognized non-helium photospheric line is the \ion{C}{iii} 4650.16 \AA\ line, which however, forms a
blend with two \ion{N}{iii} lines at 4640.64 and 4641.90 \AA. This line gives a more negative LSR radial velocity of $-$29~km\,s$^{-1}$. The accuracy
of this value is at least of the order of 15~km\,s$^{-1}$. The line can also be affected by the circumstellar material, which also ensures that  this
measurement is unreliable.

An alternative method for deriving the systemic velocity can be used if photospheric lines are not seen in the spectrum or there are no data of
sufficient quality for these lines. It employs centroid radial velocities of emission lines, such as those of \ion{Fe}{ii}. It has been successfully
tested on complex objects, such as LBVs (e.g., \citealt{hump89}) and stars that exhibit the B[e] phenomenon (e.g., \citealt{mir02}). The resulting
systemic velocity is used to determine a kinematical distance and luminosity of the object that are in general  good agreement with those found by
other methods (e.g., spectroscopic parallax, interstellar extinction versus distance relationship). The method seems to work for any type of
centrally symmetric circumstellar envelopes. Moreover, it has been shown to trace the central star motion even in binary Be stars (e.g.,
\citealt{har00}), whose disks may contain asymmetric density perturbations (see \citealt{oka91}).

\citet{dre97} comment that the centroid LSR radial velocities of the H$\alpha$, H$\beta$, and the \ion{He}{i} 5876 \AA\ lines are located between +2
and +20~km\,s$^{-1}$. Other measurements, including those we made in the optical spectra and which we have at our disposal, are listed in Table
\ref{published_optical_spectra}. The average value is $+8\pm1$~km\,s$^{-1}$, similar to $+10\pm2$~km\,s$^{-1}$ from \citet{ack08}. We  note that the
formal statistical uncertainty of these velocities may be larger, because the number of lines used is relatively small. In addition, a systematic
error due to, for instance, a slight asymmetry in the line profiles (see those shown by \citealt{zick03}) cannot be excluded. Therefore, we assume
that the real uncertainty in the systemic velocity estimated with this method might be as large as $\sim$5~km\,s$^{-1}$.

To summarize the above discussion, we conclude that the positions of the emission lines suggest a systemic radial velocity of MWC~297 of
$+8\pm5$~km\,s$^{-1}$. In contrast, the \ion{He}{i} absorption lines give $-7$~km\,s$^{-1}$ with an even  larger uncertainty, because of their
possible displacement from the photospheric position due to processing by the circumstellar material. The value derived from the emission lines seems
to be more reliable because it is based on  data from several spectra obtained over a period of 15 years. Nevertheless, it still needs to be
confirmed using high-resolution data of more lines.


\bibliographystyle{aa}

\bibliography{15676}

\end{document}